# Kuiper Belt Formation via Grainy Planetary Migration

Patryk Sofia Lykawka[1], Jonathan Horner[2], Pedro Bernardinelli[3]

[1] Kindai University, Shinkamikosaka 228-3, Higashiosaka, Osaka, 577-0813, Japan; patryksan@gmail.com
[2] University of Southern Queensland, Toowoomba, Queensland 4350, Australia
[3] DiRAC Institute, Department of Astronomy, University of Washington, 3910 15th Avenue Northeast, Seattle, WA 98195, USA

## Abstract

We used N-body simulations to model the 4.5 Gyr orbital evolution of the early Kuiper Belt, incorporating a massive protoplanetary disk, the four giant planets, and 1500 primordial Pluto-class bodies ("Plutos") that drove Neptune's grainy migration. The analysis of 67 simulated systems revealed key insights: (1) All systems featured the primary trans-Neptunian object (TNO) populations: cold/hot classical, resonant, scattered, and detached; (2) Captures into stable resonant orbits favored close Neptunian mean motion resonances (MMRs; e.g., 3:2, 2:1), while distant ones beyond 50 au (e.g., 5:2 MMR) were underpopulated; (3) Optimal matches to observed resonant fractions and the classical region (including the kernel) arose from models considering a jumping Neptune, self-gravitating Plutos, and an initial disk edge at 45–47 au; (4) Models including primordial scattered disks boosted distant MMR captures but overproduced scattered objects; (5) All models were inefficient at producing the detached ($q > 40$ au) and high-$i$ ($i > 45°$) populations and failed to populate observed niches, such as distant detached ($a > 245$ au), low-$i$ detached ($i < 20°$), low-$i$ scattered with $q = 37$–$40$ au ($i < 20°$), and extreme ($q > 50$ au or $i > 50°$) TNOs; (6) Grainy migration effects peaked early, fading as the Plutos were removed; (7) With a few primordial Plutos surviving inside 50 au, the initial population was estimated at ~150–500 to explain Pluto's solitary status. Although our four-giant-planet models reasonably replicate the trans-Neptunian structure within 50 au, they fail to account for detached, high-$i$, and extreme TNOs. Additional processes (e.g., a distant undiscovered planet) are required for a comprehensive outer solar system framework.

Key words: Kuiper belt: general; planets and satellites: formation; planets and satellites: dynamical evolution and stability; planets and satellites: general; protoplanetary discs; minor planets, asteroids: general.

*******

*******

# 1. Introduction

Trans-Neptunian objects (TNOs) orbit the Sun beyond Neptune with semimajor axes ($a$) > 30 au, within the so-called Kuiper Belt, a region that extends to approximately 2000 au (e.g. Gladman et al. 2008; Kaib et al. 2019; Gladman & Volk 2021; Nesvorny et al. 2023, and references therein). These objects are materials left over after planet formation in the outer solar system (e.g. Malhotra 1995; Levison et al. 2008; Lykawka 2012; Nesvorny 2018; Prialnik et al. 2020, and references therein) and therefore contain crucial information about the formation and orbital evolution of the giant planets, the properties of the protoplanetary disk from which the solar system formed, and the origins of several small-body populations (i.e. the Jovian or Neptunian Trojans, comets, Centaurs, and Oort cloud objects) (e.g. Morbidelli et al. 2005; Lykawka et al. 2009; Dones et al. 2015; Nesvorny 2018; Horner et al. 2020; Gladman & Volk 2021, and references therein).

The trans-Neptunian region exhibits a complex orbital structure, in which TNOs are typically considered to be grouped into four dynamical classes: resonant, classical, scattered, and detached objects (Lykawka & Mukai 2007b; Gladman et al. 2008). Although our analysis in this study is focused on the distant Kuiper Belt (at $a$ > 50 au and up to 1000 au from the Sun), we also consider additional constraints imposed by other populations located within 50 au. In addition, some TNOs exhibit puzzling orbital properties that challenge the current representative models for the formation of the outer solar system and Kuiper Belt, particularly objects with very high orbital inclinations or perihelion distances (e.g. Sheppard et al. 2016; Kaib et al. 2019; Gladman & Volk 2021; Lykawka & Ito 2023). Any appropriate model for the origin of objects in the trans-Neptunian region should explain the main properties of all of these populations. We explore this in detail below.

*Resonant TNOs* are objects locked into various Neptunian mean motion resonances (MMRs). Prominent populations include those with 3:2 ($a_{res}$ = 39.4 au), 2:1 ($a_{res}$ = 47.8 au), and 5:2 ($a_{res}$ = 55.4 au) MMRs, and possibly many other MMRs (Lykawka & Mukai 2007b; Gladman et al. 2012; Crompvoets et al. 2022). Notably, a large proportion of resonant TNOs exhibit libration within their host resonances that can persist over Gyr timescales. Such Gyr-stable resonant TNOs have been shown to be trapped within several MMRs within 100 au, particularly those with n:1 or n:2 ratios (Lykawka & Ito 2023). The only models that reproduce these populations imply that the MMRs of Neptune swept through a substantial population of disk objects during planet migration, with some models indicating the presence of a primordial scattered disk at that time (e.g. Malhotra 1995; Hahn & Malhotra 2005; Lykawka & Mukai 2007a). Conversely, the less stable resonant TNOs were established via temporary capture in MMRs. Therein, libration behaviours typically persist only for timescales ranging from thousands to several million years per capture, depending on the resonance[1]. Unlike the stable resonant population, which can be viewed as primordial, dating back to the time of solar system formation, transient resonant TNOs have been relatively recently captured and will escape those resonances again in the future. Despite individual losses, the population size remains roughly constant, as objects from classical, scattered, and detached populations can become temporarily trapped in resonances during their orbital evolution (Lykawka & Mukai 2006; Lykawka & Mukai 2007c; Yu et al. 2018).

*Classical TNOs* exhibit orbits that are concentrated within the classical region of the Kuiper Belt, historically termed the Edgeworth-Kuiper Belt following the seminal work of Edgeworth 1949 and Kuiper 1951, with $a \sim$ 40–50 au. Members (objects) of the classical population are not stably locked into a strong MMR, such as the 5:3 ($a_{res}$ = 42.3 au), the 7:4 ($a_{res}$ = 43.7 au), or the 2:1 ($a_{res}$ = 47.8 au) MMR. Classical TNOs can be further classified into dynamically cold and hot populations, which are represented by classical objects with inclinations (*i*) of < 5° and > 10° respectively (Van Laerhoven 2019; Huang et al. 2022a). Objects with intermediate inclinations ($i$ = 5–10°) may constitute a mix of the two populations. However, for simplicity, and following the literature, in this study we consider that an inclination threshold of 5° divides the two populations. Cold classical TNOs typically exhibit low eccentricities ($e \lesssim$ 0.1), but their hot counterparts exhibit a much wider range of eccentricities, up to ~0.25. Cold and hot classical TNOs also evidence distinct properties. The cold population features a higher fraction of binaries, greater albedos (more reflective), and (predominantly) very red colors. The hot population spans a wide colour range from blue to red (e.g. Gladman & Volk 2021; Bernardinelli et al. 2025). All these features strongly imply that the two populations originated in the protoplanetary disk at different locations or times. Currently, the consensus is that the cold population probably formed in situ and thus represents the remnants of the disk that existed beyond ~30 au. In contrast, the bulk of their hot counterparts likely formed within this boundary and were later transported to the classical region (by the time that the migration of Neptune concluded) (Gomes 2003; Dawson & Murray-Clay 2012; Nesvorny et al. 2020).

*Scattered* and *detached TNOs*. The orbits of scattered TNOs are more eccentric than those of other populations and span a wide range of semimajor axes[2]. Typically, the perihelia $q$ values are < 40 au, and the population was subjected to Neptune-induced scattering events during its past orbital evolution in the scattered-disk component of the Kuiper Belt. The detached TNOs move in similar orbits, but are typically of $q > 40$ au and are not locked stably into any MMR. These properties satisfy the condition that these objects exhibit decoupled (i.e. "detached") gravitational interactions with Neptune over Gyr-timescales (Lykawka & Mukai 2007b; Batygin et al. 2021). Therefore, detached TNOs generally exist in stable non-scattering orbits within the scattered disk[3] (Gladman et al. 2002; Lykawka & Mukai 2007b). Following Lykawka & Ito (2023), and in agreement with recent estimates (Beaudoin et al. 2023; Bernardinelli et al. 2025), we consider that the intrinsic ratio of scattered and detached populations in the distant Kuiper Belt is ≤ 1. Notably, resonant or von Zeipel-Lidov-Kozai (vZLK) interactions (e.g. Ito & Ohtsuka 2019), or chaotic diffusion, cannot explain the detached population, in particular those members for which $i < 20°$ or $a > 150–250$ au (Gomes et al. 2005; Lykawka & Mukai 2007c; Sheppard et al. 2016; Nesvorny et al. 2016; Kaib & Sheppard 2016; Pike & Lawler 2017; Gladman & Volk 2021; Hadden & Tremaine 2024; Kaib et al. 2025). Therefore, there is ongoing debate as to whether external perturbations—beyond those induced by the currently known planets—are required to explain the orbital variety and the prominent population of detached TNOs (Huang et al. 2022b; Lykawka & Ito 2023; Nesvorny et al. 2023; Kaib et al. 2025).

---

[1] Such "temporary capture" seems to be a regular evolutionary feature of unstable solar system objects (e.g. Horner & Evans 2006). Several examples of such captured objects have been identified in the resonant populations of the solar system (e.g. Horner & Lykawka 2012; Hui et al. 2021, 2024; Carruba et al. 2025; Greenstreet et al. 2025).

A high-$i$ reservoir of TNOs has been suggested in the literature (Lykawka & Mukai 2007b; Chen et al. 2016; Kaib et al. 2019; Gladman & Volk 2021; Kaib & Volk 2024). In this context, a *high-i TNO* is an object with $i > 45°$. Such an intriguing subpopulation cannot be explained by any of the several existing outer solar system scenarios, including the standard Kuiper Belt model of giant-planet instability followed by grainy outward migration of Neptune[4] (Lykawka & Mukai 2007c; Gomes et al. 2008; Lykawka & Mukai 2008; Nesvorny & Vokrouhlicky 2016; Nesvorny et al. 2016; Kaib & Sheppard 2016; Sheppard et al. 2016; Pike et al. 2017; Nesvorny 2018; Kaib et al. 2019; Gladman & Volk 2021). Such models produce populations in which only an intrinsic < 2% of objects belong to the high-$i$ class, in conflict with observations. In other words, using the orbital information of the AstDyS observational database[5], the apparent fraction of distant TNOs ($a > 50$ au) moving on highly inclined orbits is 2% (3%) for $q > 25$ au ($q > 15$ au). Notably, these fractions were derived from a population with observational biases that disfavor the discovery of higher-inclination TNOs (Gladman et al. 2012; Chen et al. 2016; Kavelaars et al. 2020; Gladman & Volk 2021), implying that the intrinsic population fraction is likely much higher than that observed. Thus, the origin of the high-$i$ population probably requires additional perturbations beyond those imparted by known planets of the solar system, with potential candidates including undiscovered distant planets or early stellar encounters (Kaib et al. 2019; Nesvorny et al. 2023; Lykawka & Ito 2023). Here, we conservatively assume that the intrinsic TNO high-$i$ fraction is at least 2% in general, or at least 1% and 7% within the scattered and detached populations, respectively.

---

[2] We note that there may be certain transition zones with no clear boundaries between hot classical and scattered TNOs in $a$–$e$ space. For example, scattered TNOs of $q \sim 35$–40 au in orbits close to the classical region could be viewed as part of the hot classical population, and vice versa. Given the scope of this paper, we adopt a simple classification consistent with prior studies, because more complex schemes would not alter our principal results (see Footnote 3 and Section 2).

[3] We verified that TNOs that interact either strongly or weakly with Neptune tend to evolve with $q < 35$ au and $q \sim 35$–40 au, respectively. Although the former and latter populations could be classified as "scattering" and "detached" respectively (e.g. see Gladman et al. 2008 and Nesvorny et al. 2023), we use the term "scattered" for the entire population with $q \sim 25$–40 au, for consistency and simplicity. We also confirmed that our results do not require adaptation of the classification scheme of Gladman et al., or similar schemes that define scattered and detached TNOs. Nor was it necessary to consider an alternative $q \sim 15$–40 au range for the scattered population.

[4] Henceforth, for brevity, "instability" and "migration" will refer to events related to the giant planets.

Another puzzling subpopulation is represented by *extreme TNOs*, which are members of detached or high-$i$ reservoirs that possess peculiar orbits that stand out within their parent populations (e.g. Sedna has a $q \sim 76$ au). The existence of extreme TNOs represents an additional useful constraint that can indicate the specific mechanisms that played important roles in the history of the solar system (Brown et al. 2004; Trujillo & Sheppard 2014; Sheppard & Trujillo 2016; Sheppard et al. 2019; Chen et al. 2025; Kaib et al. 2025). As discussed in Lykawka & Ito (2023), several models, some of which also include external perturbations, and typical dynamical mechanisms, such as coupled MMR+vZLK interactions during or after Neptune's migration, are unlikely to explain the orbits of extreme TNOs. As of July 2025, orbital information retrieved from the AstDyS database allowed us to identify 15 extreme TNOs (Table 1). The orbital diversity of these TNOs poses a remarkable challenge to any model that aims to reproduce the orbital structure in the trans-Neptunian region. Clearly, more discoveries and follow-up observations during current and future large-scale surveys of the trans-Neptunian region are required to characterize the populations discussed above better, in particular the underrepresented detached, high-$i$, and extreme TNOs (Petit et al. 2011; Alexandersen et al. 2016; Petit et al. 2017; Bannister et al. 2018; Ivezic et al. 2019; Bernardinelli et al. 2022; Trilling et al. 2024; Kurlander et al. 2025). The orbital distribution of TNOs and their main dynamical classes are summarized in Figure 1.

---

[5] The AstDyS observational database of TNOs is publicly available at https://newton.spacedys.com/astdys/
We selected only observed TNOs, the best-fit orbits of which contained small orbital uncertainties (i.e. the 1-sigma uncertainties of both the semimajor axis and eccentricity were within 1% of the nominal orbital elements).

Earlier models of Kuiper Belt formation focused on specific populations such as the resonant TNOs (Schulz 2002; Hahn & Malhotra 2005; Morbidelli et al. 2008; Lykawka 2012). The central premise of such models is that the four giant planets initially formed in a spatially compact configuration and later migrated through the protoplanetary disk because they interacted with disk objects during the early history of the solar system (Fernandez & Ip 1984; Nesvorny 2018). Neptune's outward migration likely played a fundamental role in shaping the Kuiper Belt, as the planet gravitationally scattered a substantial population of the disk, and many MMRs then swept the disk in lockstep. This would be expected to have populated various MMRs via resonant capture during migration (Hahn & Malhotra 2005; Gladman et al. 2012; Pike & Lawler 2017). However, one notable issue with such models is that the generated resonant population is far larger than that observed (this is the so-called resonance overpopulation problem) (Nesvorny 2015a; Nesvorny & Vokrouhlicky 2016). Similar models sought to reproduce the main properties of the four giant planets by considering a primordial outer solar system packed with several planets that experienced an instability (Tsiganis et al. 2005; Nesvorny 2011; Batygin et al. 2012; Nesvorny 2015b; Nesvorny 2018; Clement et al. 2021). Notably, the Nice model (e.g. Nesvorny 2018) addresses Kuiper Belt formation in the context of an instability. However, as discussed in previous studies, this model probably cannot explain certain key properties of the Kuiper Belt, such as the cold classical component, or the main constraints of the distant Kuiper Belt (Gladman et al. 2012; Pike et al. 2017; Lawler et al. 2018; Lawler et al. 2019; Volk & Malhotra 2019; Lykawka & Ito 2023).

More recent models of Kuiper Belt formation consider a five-giant planet instability that is followed by residual, grainy, and slow planet migration (Nesvorny & Vokrouhlicky 2016; Kaib & Sheppard 2016; Nesvorny et al. 2020; Kaib et al. 2024; Kaib et al. 2025). In such models, the migration of Neptune is associated with minor erratic variations in the semimajor axis of the planet as the planet evolves outward. Such grainy behaviour may solve the resonance overpopulation problem by reducing the capture and retention of objects in MMRs. Such models frequently imply that Neptune experienced an orbital discontinuity during migration attributable to the gravitational scattering of an additional planet that was later lost from the solar system during the instability. Such a "jumping Neptune" event could explain the concentration of a group of cold classical TNOs around $a$ = 43.8–44.4 au, the so-called kernel of the Kuiper Belt (e.g. Petit et al. 2011; Nesvorny 2015b). A comprehensive version of the model also considered galactic tides and the effects of early passing stars, which were invoked to explain the properties of some extreme TNOs such as Sedna (Nesvorny et al. 2023). Therefore, it is thought that a jumping Neptune followed by grainy migration is currently the optimal model for Kuiper Belt formation; this is henceforth termed the "fiducial" model. However, as discussed above and elsewhere (Lykawka & Ito 2023), no existing model can consistently satisfy all of the Kuiper Belt constraints. Therefore, further investigations are warranted to understand the orbital evolution and structure of the Kuiper Belt better.

Models involving grainy migration and similar scenarios propose that the protoplanetary disk of the outer solar system featured a size distribution of objects, several of which were Pluto-class objects, with a few Earth-class objects, i.e. bodies with masses comparable to those of Pluto/Eris and Earth, respectively (Nesvorny & Vokrouhlicky 2016; Shannon & Dawson 2018). The grainy migration model requires the presence of hundreds or thousands of Pluto-class objects (termed "Plutos" below for brevity) in the primordial disk at the time of migration (Nesvorny & Vokrouhlicky 2016; Kaib & Sheppard 2016; Kaib et al. 2024; Kaib et al. 2025). Pluto, Eris, and possibly other undiscovered dwarf planets located beyond 50 au represent the current remnants of this primordial population. By the same token, it remains plausible that an Earth-class planet acquired a stable orbit in the distant Kuiper Belt, supporting the hypothesis of a resident undiscovered planet within the solar system (Eriksson et al. 2018; Bailey & Fabrycky 2019; Izidoro et al. 2025). The long-term perturbations of such a hypothetical Kuiper Belt planet ("Planet X") could shape the trans-Neptunian region over Gyr, thus potentially explaining the distinct populations and peculiar properties of that region (Lykawka & Mukai 2008; Lawler et al. 2017; Lykawka & Ito 2023). However, the origin of the distant and stable orbit of the putative planet is still not well understood. Also, this scenario has yet to consider the influence of planetary migration, account for the inner Kuiper Belt (at $a < 50$ au), or better constrain the high-$i$ component using higher-statistics observational data.

In this study, building on the preceding discussion, we aim to address the main constraints apparent in the Kuiper Belt, with a particular focus on models that incorporate giant planet residual grainy migration. Therefore, models involving temporary "rogue" planets (Gladman & Chan 2006; Silsbee & Tremaine 2018; Huang et al. 2022b), hypothetical sub-Earth/Earth-class (Lykawka & Mukai 2008; Lykawka & Ito 2023; Siraj et al. 2025) to super-Earth-class planets (Sheppard & Trujillo 2016; Batygin & Brown 2016), disk secular instabilities (Madigan & McCourt 2016), or Oort cloud formation (Kaib et al. 2011; Brasser & Schwamb 2015) during the early solar system are not discussed here. Specifically, our investigation has the following features: residual grainy migration, driven by a massive protoplanetary disk with several thousand embedded objects and primordial Plutos; gravitational influence of the primordial Plutos on the planets (no artificial forces or analytical prescriptions are used); and a fiducial model of the solar system based on the jumping Neptune scenario and grainy migration. The details are discussed in the following section.

## 2. Methods

We used N-body simulations to investigate the formation and dynamical evolution of the Kuiper Belt. The four giant planets were considered throughout all investigations described in this study. Our models included giant planet residual grainy migration induced by massive planetesimals and Plutos in the protoplanetary disk, as described below. In all our simulations, the four known giant planets commenced on low-$e/i$ (i.e. stable) orbits, and with the Jupiter-Saturn pair located in a manner ensuring that their ratio of orbital periods PS/PJ was greater than 2. All planets interacted with the massive objects in the system via exchange of energy and angular momentum, causing the giant planets to migrate through the disk as a result of these interactions. Jupiter, Saturn, Uranus, and Neptune were placed at 5.26 au, 9.2 au, 18.7 au, and 24.0 au at the start of all simulations. The four giant planets commenced with eccentricities of 0.07, 0.07, 0.01, and 0.01 and inclinations of 2°, 2.5°, 1.5°, and 1.5°, respectively. In the jumping Neptune scenario, we also tested an initial $e = 0.075$ for Neptune (see below)[6].

All simulations considered a disk consisting of two main components: an inner massive disk extending from 25 au to 30 au and an outer massless disk starting at 30 au and continuing to 49 au. To improve the number statistics of our final systems, we also included a massless inner-disk component. In each run, the massive and massless inner disks had approximately 13000~16000 and 87000~140000 objects (so the inner disk contained ~100000-156000 objects in total), and the outer disk contained ~1800-2500 objects. The inner (outer) disk objects initially exhibited eccentricities and inclinations within 0.2 (0.01) and 5° (0.5°), respectively. The more dynamically excited state of the inner disk would be expected to result from gravitational perturbations induced by the giant planets, Plutos, and other planetary bodies during the early stages of evolution, presumably both before and during the instability. In contrast, the outer disk would be expected to have been dynamically cold during this early period (Van Laerhoven et al. 2019). The initial conditions described above are similar to those employed by several past representative models (Nesvorny 2015a; Nesvorny & Vokrouhlicky 2016; Kaib & Sheppard 2016; Volk & Malhotra 2019; Nesvorny et al. 2020). Inspired by some of these models, we placed 1500 Plutos within the massive component of the inner disk in all systems. By default, all massive disk objects fully interacted with the planets in the system, but not with one another. In our simulations, in order for Neptune to attain a final orbit close to its current position after 4.5 Gyr of evolution, the massive disk was required to contain ~14.5 Earth masses (ME) in total, with approximately 5 ME of that mass represented by the Plutos[7] (e.g. consistent with Nesvorny & Morbidelli 2012; Nesvorny 2015a; Nesvorny 2018). As it is thought that the inner disk was roughly three orders of magnitude more massive than the outer disk in the early solar system (e.g. Nesvorny et al. 2020), modelling of the latter using only massless particles was appropriate for our purposes. The above conditions allowed our planets and disk objects to interact mutually so that the giant planets migrated in a gradual manner to their current orbits. Nesvorny & Vokrouhlicky (2016) and Kaib & Sheppard (2016) mimicked grainy migration by adding an analytical setup to their simulations. However, in our modelling, grainy migration occurred in a self-consistent way because our Plutos interacted directly with Neptune (as in Kaib et al. 2024). In a complementary setup, we also tested systems in which the Plutos gravitationally interacted with one another. Finally, we added primordial scattered objects to some systems to test their effects on the results, particularly any possible contribution to resonant populations in distant MMRs beyond 50 au (e.g. Gladman et al. 2012). Such primordial scattered disks might have existed during or after the instability. For each combination of the initial conditions of the giant planets and the disk in a particular scenario, we performed at least 10 runs using distinct random seeds. Neither galactic tides nor passing stars were included in the modelling because their effects typically play roles at only $a > 1000$~$2000$ au (Gladman et al. 2008; Nesvorny et al. 2017; Kaib et al. 2019; Saillenfest 2020; Gladman & Volk 2021; Kaib & Volk 2024).

---

[6] The instability can produce diverse orbital evolutions of the giant planets from initial five- or six-planet configurations (e.g. Batygin et al. 2012; Nesvorný & Morbidelli 2012; Clement et al. 2021; Kaib et al. 2024). For the purposes of this study, we simplified the instability stage to an idealized relatively weak event, thus focusing on successful cases described in previous studies. Although this limited possible insights into the full range of instability effects, it better isolated the influence of the fundamental initial conditions in our distinct models, imposing useful constraints on the giant planet and disk conditions during Kuiper Belt formation.

The fiducial model featured the initial conditions of the standard model described above, but also included an orbital evolution akin to the jumping Neptune scenario. As in previous studies (Nesvorny 2015b; Nesvorny & Vokrouhlicky 2016; Kaib & Sheppard 2016; Kaib et al. 2019; Nesvorny et al. 2020), we stopped the simulations when Neptune attained $a \sim 28$ au, and then restarted them after applying $\Delta a = +0.5$ au to the recorded $a$ and setting $e = 0.075$ for the planet ($\Delta e = +0.05$–$0.07$). This relatively small eccentricity of the post-jump Neptune is required to ensure that the cold classical population exhibits properties compatible with those observed (Wolff et al. 2012; Dawson & Murray-Clay 2012; Nesvorny 2021). However, unlike in previous studies, both the pre-jump and post-jump Neptune here underwent grainy migration through the planetesimal disk attributable to mutual interactions with massive objects and Plutos.

---

[7] Note that in the fiducial (jumping Neptune) model, the initial mass in the massive disk had to be reduced to 12.5 ME, so that the post-jump Neptune could acquire its current orbit by the end of the simulations. In addition, Pluto-class objects had individual masses of 0.00333 ME, comparable to those of Pluto (i.e. 0.00333/0.00218. ~ 1.5-fold), Eris (~1.2-fold), and Triton (~0.9-fold).

We investigated the long-term dynamical evolution of all systems by executing simulations up to 4.5 Gyr, thus covering the age of the solar system. We used a modified version of the MERCURY integrator to conduct the simulations (Chambers 1999; Hahn & Malhotra 2005). The simulation time step was 0.6 yr. Objects that evolved to heliocentric distances of less than 3 au or greater than 1000 au were discarded from the simulations. Table 2 summarizes the initial conditions of the models considered in this work. All simulation runs were executed using single CPU cores of the general-purpose PC cluster of the National Astronomical Observatory of Japan, and personal multi-core workstations. Although most individual runs required ~3–4 months to finish, those employing self-gravitating Plutos took approximately five times longer (i.e. ~15–20 months).

After 4.5 Gyr of orbital evolution, the individual Neptune analogues evolved quite similarly and acquired final orbits between 29 and 31 au. Specifically, in 50 of 67 systems utilized in our analysis, the values fell roughly within ± 0.5 au of the Neptune $a$ of 30.1 au. We normalised the orbits of the objects in all systems (runs) by applying a scaling factor of $30.1 \times a_{\text{run}} / a_{\text{Nep,run}}$, where $a_{\text{run}}$ and $a_{\text{Nep,run}}$ refer to the final semimajor axis of an object and the Neptune analogue, respectively. Next, we rescaled the associated perihelia of the disk objects. Given the similar orbital behaviour of our Neptune analogues, and the fact that ~90% (67 of 75) of the resulting Kuiper Belts contained the four main populations discussed in Section 1, we combined the simulation results of the runs within each model. Thus, we could effectively probe individual models at high resolution. This procedure also enabled a statistical analysis of the results by mitigating the stochasticity inherent in the chaotic evolution of single systems (e.g. Volk & Malhotra 2019).

Finally, we classified model objects using the algorithm described below, based on the final orbital elements and the defining criteria for the following dynamical classes and key subpopulations. 1) *Resonant*: We used our RESTICK code to calculate the libration angles of MMRs during the orbital evolution of each object (Lykawka & Mukai 2007a). Resonant objects exhibited libration for a total duration 2–4.5 Gyr, during orbital evolution. Therefore, our identified resonant population consisted of stable resonant objects. Conversely, the transient resonant counterparts were identified as members of other populations, particularly those that were scattered or detached[8]. 2) *Classical*: Non-resonant objects located approximately between MMRs 3:2 and 2:1 (i.e. at $a$ = 40.0–47.3 au, outside the resonance boundaries) with $q > 35$ au. In addition, we included a few objects located beyond 2:1 MMR (up to $a \sim 60$ au) and with $q > 40$ au that originated beyond 30 au in the outer region of the protoplanetary disk. The cold and hot classical components consisted of classical objects with $i < 5°$ and $i > 5°$ respectively. 3) *Scattered*: Non-classical objects with $25 < q < 40$ au. 4) *Detached*: Non-classical objects with $q > 40$ au. For completeness: 5) *Centaurs*: Objects with $q < 25$ au. The identified scattered, detached, and Centaur objects were not stably resonant. Finally, we turned to some peculiar subpopulations. High-$i$ objects exhibited $i > 45°$. We defined objects with $q > 60$ au, $i > 60°$, $q > 50$ au ($i < 20°$) or $i > 50°$ ($q < 40$ au) as extreme objects (Lykawka & Ito 2023). Given that our focus was on the distant Kuiper Belt, and as observational data are currently limited, the results do not significantly depend on the classification details (Section 1).

---

[8] Lykawka & Mukai (2007c) and Yu et al. (2018) found that objects scattered by Neptune over Gyr-long timescales spent ~40% of their orbital evolution temporarily captured in MMRs in the scattered disk.

## 3. Results and Discussion

As planets interacted with disk objects, as described previously, Neptune migrated outward in a grainy manner, sweeping the inner region of the protoplanetary disk until the planet acquired its final orbit. Overall, only a small proportion of inner-disk objects survived our simulations. In contrast, a larger fraction of outer-disk objects survived because they were less affected by the migrating Neptune. Our classification algorithm appropriately divided the final system objects into the main Kuiper Belt dynamical classes (Section 1). In the investigation below, we focus on these populations and how their formation depends on certain key factors of our five models. We therefore excluded individual systems exhibiting any of the following observation-contradicting features: apparently non-existent resonant populations within 50 au, absence of 3:2 MMR members with $e > 0.15$–$0.2$ ($q < 33.5$–$31.5$ au), or excessive dynamical stirring/depletion of the cold classical region. In these systems, Uranus and Neptune analogues either evolved to an orbital period ratio $PN/PU \geq 1.96$ or underwent stochastic perturbations exciting the eccentricity of Neptune to 0.04–0.06, or both. Such behaviours strengthened the near-2:1 MMR between Uranus and Neptune, the overlapping Neptunian MMRs at high eccentricities, and the perturbations of secular resonances in the Kuiper Belt (e.g., secondary resonances), all of which can cause destabilisation within 50 au (e.g. Dawson & Murray-Clay 2012; Wolff et al. 2012; Volk & Malhotra 2019; Nesvorny et al. 2020; Graham & Volk, 2024; Kaib et al. 2024). Ultimately, after passing our obtained systems through the filter described above, 67 were deemed acceptable for the purposes of this study. Thus, ~90% (67/75) of the systems (runs) were utilized to derive the results that we present.

### 3.1. Orbital evolution of Neptune

Neptune underwent orbital migration during the entire 4.5 Gyr of dynamical evolution, a process sustained by the slowly decaying population of massive objects in the disk. At late stages, the planet acquired a near-circular, low-$i$ orbit near 30 au. In general, and consistent with previous studies (Nesvorny & Vokrouhlicky 2016; Kaib et al. 2024), Neptune's migration was fast during the first Myr of evolution and then slowed, becoming negligible after ~1 Gyr. Figure 2 shows an example of this behaviour, highlighting the orbital evolution of Neptune in the 15 runs of the fiducial model (grainy migration with jumping Neptune). The obtained Neptune analogues confirmed these migrational trends and also exhibited eccentricity damping during the first tens of Myr because of dynamical friction imparted by massive objects in the disk. The final orbital configurations acquired by these Neptune analogues were remarkably similar to that of the real Neptune (average $a_N$ = 30.1 au and $e_N$ = 0.006; standard deviations of 0.2 au and 0.002). We confirmed that the Neptune analogues of the other models exhibited similar orbital behaviours and final properties. Therefore, we conclude that, for our present purposes, the migration of Neptune was adequately characterized in our simulations.

When analysing the migration behaviours of the Neptune analogues of all systems across the models, the late orbital evolution (i.e., $a_N$ > 28 au) implied that migration was better described by a decay law more complex than that of a single exponential decay. In particular, we verified that a double or stretched exponential, or a power law, acceptably fitted the results, remaining compatible with the fast early orbital evolution ($a_N$ < 28 au). Kaib et al. (2024) also noted the poor fit of a single exponential function in scenarios similar to our 4GP_jNep and 4GP_pluSG models (see Table 2). For brevity and to facilitate comparison with previous studies, we fitted the orbital evolution of our Neptune analogues to a double exponential function of the form exp(-$t$/tau1) + exp(-$t$/tau2). The typical tau1 and tau2 timescales were best represented by medians of 4 Myr and 55 Myr, respectively. This finding implies that, despite the initial rapid outward migration of the Neptune analogues, their final orbits stabilized after several hundred Myr of slower subsequent migration (a representative case is illustrated in Figure 3). Again, these results are consistent with those of Kaib et al. (2024). Thus, our simulations imply that Neptune's orbit decayed rapidly because of heavy initial scattering of disk objects (and Plutos) during early orbital evolution, and then transitioned to a gradual and smoother orbital migration as the number of massive objects decreased with time. We conclude that although the late orbital evolution of Neptune was largely insensitive to the disk property details, such evolution followed a multi-timescale process that a single exponential function cannot describe.

3.2. Dynamical evolution and survival of Pluto-class objects ("Plutos")

As expected, the interactions between Plutos in the inner disk and the giant planets rendered migration of the latter grainy over time. Indeed, all the Plutos underwent gravitational encounters with Neptune at some point in their evolution, which scattered them across the system as Neptune migrated outward. As a result, the number of Plutos decreased gradually over time, as did the number of their smaller-mass counterparts. Ultimately, only a small proportion of the initial 1500 Plutos survived after 4.5 Gyr (Figure 4). In our fiducial model, a median of eight Plutos survived, of which five were located within $a < 50$ au (Table 3). The other models yielded similar numbers of survivor Plutos, with stochastic variation on the order of factor 2. Overall, we found that the trapping efficiency of Plutos beyond 30 au was roughly 0.5%, which is about five-fold larger than suggested by Nesvorny & Vokrouhlicky (2016). Currently, Pluto and Eris are the only known Pluto-class objects in the solar system beyond the orbit of Neptune[9]. Thus, our results predict the existence of ~4–11-fold more Plutos within 30–50 au (where only Pluto presently exists) or ~3–7-fold more such objects within 30–100 au (where only Pluto and Eris now exist). These results imply that either the number of initial Pluto-class objects in the inner disk should be reduced proportionally, or that several Pluto-class objects await discovery within this region. However, if such Plutos did exist within 100 au, as indicated by these results, the surveys conducted to date should have already discovered them. Given that observational surveys are near-complete for this region (Jewitt et al. 1998; Trujillo et al. 2001; Brown et al. 2004, 2005; Schwamb et al. 2010; Sheppard et al. 2011; Schwamb et al. 2014; Kavelaars et al. 2020; Holman et al. 2025), our results imply that the number of primordial Plutos must have been between 150 and 500, which aligns with the findings of Kaib et al. (2024) and Shannon & Dawson (2018), who suggested numbers of no more than 1000. If Triton is considered to be a known Pluto-class object, our results imply that there were ~250–750 primordial Plutos. Our Kuiper Belt analogues also indicate that Plutos may have survived on any dynamical class beyond Neptune (Figures 5 and 6), typically within or near the classical region. The fact that only Pluto and Eris (and perhaps Triton) have been found within a similar orbital region reinforces the inference that the number of primordial Plutos was no more than 500 (750).

3.2.1. Note on the origin of the orbits of Pluto and Eris

We used our data to explore whether Pluto-class objects that survived in the 67 final systems could acquire orbits similar to those of Pluto or Eris. In this context, we assumed that a Pluto-like orbit was a 3:2 resonant orbit, irrespective of eccentricity or inclination, whereas an Eris-like orbit required that $a > 50$ au, $q = 35–45$ au, and $i > 40°$. For simplicity, and to improve the statistics, we combined the results from models that did not include self-gravitating Plutos (58 systems in total) and compared these data to those of the model that included this effect (nine systems from model 4GP_pluSG). At the end of the simulations, we found that the probability that a primordial Pluto-class object acquired a 3:2 resonant orbit was ~0.1% (i.e. 85 objects of an initial 87000) in non-self-gravitating models and ~0.02% (3 of an initial 13500) in the self-gravitating model. The probabilities of Eris-like orbits were ~0.005% (4 of 87000) and ~0.01% (2 of 13500), respectively. Thus, the more realistic 4GP_pluSG model could reproduce the orbits of Eris and Pluto with probabilities of ~15% and ~30%, respectively (i.e. 1500 primordial Plutos × ~0.0001[2]). However, these probabilities are lower if no more than 1000 such objects existed in the inner region of the protoplanetary disk. This is of concern. In summary, the origins of the orbits of Pluto and Eris remain challenging to explain. Further investigations are warranted.

---

[9] Neptune's moon Triton might be an additional member of the primordial Pluto population, with some models indicating that Triton was once part of a binary Pluto pair captured and disrupted by a close encounter with the giant planet (see Section 4.1.3 of Horner et al. 2020, and references therein). Similarly, other Pluto-class objects could remain undetected if they are on distant orbits in the scattered and detached populations.

### 3.3. Influence of self-gravitating Plutos

We investigated the influence of self-gravity among the Plutos by comparing the results of the 4GP and 4GP_pluSG models. First, gravitational interactions among Plutos, especially in the early dynamical stages when their numbers were highest, substantially affected the retention of captured resonant objects. In detail, we found that the proportions of inner-disk objects that ended up in the resonant population were approximately 41% and 15% in the 4GP and 4GP_pluSG models, respectively. In short, additional perturbations from Plutos on resonant objects destabilized libration (i.e. resonant behaviour ceased) in a significant fraction of the resonant population, in agreement with the results of Kaib et al. (2024). Although such disruption was more pronounced during the first tens of Myr of system evolution, when primordial Plutos were more numerous than later, perturbations likely persisted throughout system evolution via continuous gravitational encounters between Plutos and resonant objects. As anticipated, inner-disk objects that failed to enter resonant orbits became parts of other populations. In particular, notable increases in the proportions of hot classical (from 12% to 20%), scattered (from 37% to 47%), and detached (from 7% to 12%) populations occurred when the self-gravity of Plutos was included in the assessments. We observed similar trends when examining the final populations of combined inner- and outer-disk objects across all systems (Table 3; see also Table A1 in the Appendix[10]). The 4GP_pluSG model revealed more pronounced resonance depletion within ~50 au compared to the region beyond it. This occurred because Plutos remained on orbits of a few tens of au for prolonged periods while interacting with resonant populations during the migration of Neptune.

The perturbations caused by the Plutos affected the local population that originally formed beyond 40 au. That population is the likely primary source of the cold classical Kuiper Belt. To isolate the influence of Neptune's scattering and the resonance effects of the 5:3 and 2:1 MMRs, we compared the final orbital distributions of local (outer-disk) objects that were classified as cold classicals within $a$ = 42.7–47.3 au. Although the semimajor axes were broadly similar across the models, the 4GP_pluSG model produced a cold classical population with higher median inclinations (3.6° vs. 1.8–2.2°) and eccentricities (0.075 vs. 0.055–0.060) than those of the other models. Consequently, the cold classical objects acquired dynamically hotter and more dispersed orbits attributable to additional stirring by the self-gravitating population of Plutos (Figure 7). In summary, given that the orbital distributions of classical TNOs are crucial when seeking to understand the coexistence of cold and hot classical populations, future studies incorporating mutual perturbations among primordial Plutos could impose valuable new constraints on the origin and dynamical evolution of both populations. Finally, the inclusion of self-gravitating Plutos slightly stirred the orbits of the overall surviving population (all dynamical classes combined), shifting the perihelia upward by ~1 au and the inclinations by ~1°.

---

[10] The results in Table 3 are those obtained when a protoplanetary disk outer component extends to 45 au. Complementary results for 47 au are summarized in Table A1.

In short, our analysis revealed that treating Plutos as fully gravitationally interacting objects yielded important effects.

3.4. Orbital structure in the Kuiper Belt

We explored whether the final systems of our models reproduced the distinct TNO populations and their primary orbital properties. Overall, these systems contained the four giant planets and the main Kuiper Belt components on orbits resembling those that are observed, namely a concentration of objects within 50 au, a broad resonant population, and an extended scattered disk. Figures 5 and 6 illustrate certain Kuiper Belt analogues that contain the classical (including the cold and hot components), resonant, scattered, and detached dynamical classes. Given the similar results within a given model, and to mitigate chaotic variability, we combined the systems (runs) across each model in the subsequent analyses. Given the complex orbital structure of the classical region and the scope of this study, a more rigorous discussion of the classical populations is reserved for the future. In the analysis below, we consider all objects across our models that survived with a final $q$ > 25 au[11]. Considering the cold classical population and the constraints imposed by the resonant population ratios (such as the 3:2/2:1 and 5:2/2:1 ratios), we found that inclusion of outer-disk objects that initially formed within 45 au or 47 au in the analysis yielded acceptable results.

---

[11] We did not consider objects with $q$ = 15–25 au in this investigation because other mechanisms not modelled here (e.g. galactic tides and passing stars) could generate objects with such orbits.

### 3.4.1. Resonant populations and the resonance overpopulation problem

We begin our analysis by noting that the systems produced prominent and stable resonant populations that represented ~12–38% of the trans-Neptunian region (Table 3). We identified Gyr-stable resonant objects in the 1:1, 5:4, 4:3, 3:2, 5:3, 7:4, and 2:1 MMRs (Figures 5 and 6). Stable resonant members were also commonly identified in more distant MMRs, such as the 5:2, 3:1, 4:1, 5:1, 6:1, and 7:1 MMRs. However, these resonances were in general less populated than were their closer counterparts. Models 4GP+SD and 4GP+hotSD, which initially included a primordial scattered disk, produced larger resonant populations in the 2:1 and other distant MMRs. These stable resonant objects can be compared to the stable resonant TNOs observed in the Kuiper Belt; the latter are likely primordial. For example, many 4-Gyr-stable resonant TNOs in the 2:1 and 5:2 MMRs exhibited libration amplitudes lower than those expected from resonance sticking alone, and distinct inclination and perihelial distributions; this indicates that they were captured in MMRs when Neptune engaged in resonance sweeping of the protoplanetary disk (Yu et al. 2018; Lykawka & Ito 2023). Our results imply that the outer disk (beyond ~38 au) must be invoked to explain the 2:1 MMR population. However, our results imply that an outer disk extending beyond 47 au is less consistent with the observational constraints. In this scenario, the 5:2 MMR would sweep an empty region beyond 47 au during migration. Thus, models incorporating a primordial scattered disk offer an additional possible source of the 5:2 MMR population. Inclusion of a primordial scattered disk could also account for the 3:1, 4:1, and 5:1 resonant populations, as these resonances likely captured more objects throughout the ~50–70 au region as Neptune migrated (Lykawka & Ito 2023).

Although the modelled resonant populations resemble those observed in typical MMRs such as the 3:2 and 2:1 MMRs (e.g. based on their relative populations), distant MMRs are generally underpopulated relative to the observed data. Specifically, our simulations produced a 3:2-to-5:2 population ratio ~1.5–35.5-fold larger than the observed intrinsic ratio of 0.9–2.0 (Table 3). Indeed, only the 4GP+hotSD model yielded a marginal ratio of 3.0. In addition, the inclination dispersion of our 3:2 resonant population is likely dynamically colder than the observed intrinsic value. Given that this trend holds across all tested models, reconciliation with observations requires either a more excited initial disk inclination distribution than assumed here (e.g. as employed in Nesvorny 2015a, Nesvorny & Vokrouhlicky 2016, and Nesvorny et al. 2020) or additional perturbations beyond those imparted by the giant planets and Plutos.

Can grainy migration of Neptune solve the resonance overpopulation problem? Overall, the results of the fiducial and the standard 4GP models imply that the answer is no[12]. Although the intrinsic fraction of observed resonant TNOs would be expected to represent 13–20% of the entire TNO population, these models predict proportions that are 1.5–2-fold larger (Table 3). In other words, resonant populations are overly abundant. On the other hand, incorporation of Pluto self-gravity into the 4GP_pluSG model substantially impacted resonance capture and retention. In this model, the resonant population fraction was ~14% (~12%) when the disk outer region extended to 45 (47 au). Furthermore, as this model was associated with more substantial depletion of inner MMRs, such as the 3:2, the model may reduce the ratio of the 3:2 to 5:2 resonant populations, which is currently a challenging constraint in the Kuiper Belt. Finally, the additional perturbation from the Plutos was associated with slightly more excited inclination distributions in the 3:2 MMR. These results were in better agreement with the intrinsic values estimated from observations than were those of other models. However, as discussed in Section 3.2, the possibility that the number of primordial Plutos was in fact smaller (~150–500) indicates that follow-up studies are needed to confirm these results. If there were fewer Plutos, any influence thereof on the migration of Neptune would be weaker, likely associated with markedly less grainy migration in the absence of other mechanisms. Second, as illustrated in Figure 4, the rapid decline in the number of Plutos implies that such weakened grainy migration would operate more effectively only during the first Myr of Neptune's orbital evolution. Although Kaib et al. (2024) found that 200 self-gravitating Plutos substantially depleted the 3:2 MMR by the time of the instability epoch, our study highlights the importance of determining the proportions of resonant populations relative to those of the entire surviving population, and the ratios of resonant populations after 4.5 Gyr of post-instability evolution [13]. Therefore, although our results imply that Neptune's grainy migration and perturbations from 1500 self-gravitating primordial Plutos may indeed resolve the resonance overpopulation problem, more detailed future investigations of disks containing only a few hundred of such Plutos are required to reach definitive conclusions.

---

[12] The 4GP+SD and 4GP+hotSD models are not considered in the context of the resonance overpopulation problem because they yielded final systems with excessively abundant scattered populations, thus artificially reducing the relative fraction of the entire resonant population.

[13] Note that Kaib et al. (2024) primarily examined the occupancy of the 5:4, 4:3, and 3:2 MMRs prior to and during the instability phase, whereas this study focuses on the final, post-instability, Gyr-stable resonant populations and the associated resonance overpopulation problem. Additionally, the results of the cited work were based on individual simulation runs. This means that the properties of the final populations may have exhibited stochastic variations (e.g. as shown in Table 4 of that study).

### 3.4.2. Classical populations and the kernel

Turning to the classical region, the cold and hot classical populations of many systems exhibited orbital properties similar to those that are observed (Figures 1, 5, and 6). As expected, the bulk of both populations originated from the outer (i.e. local) and inner regions, respectively, of the protoplanetary disk. We extended the outer disk to 45 or 47 au, both to mimic the rapid decrease in the number density of objects in the outer skirts of the classical region (the so-called Kuiper Cliff) and to investigate the formation of the kernel feature. Both are well-established observational constraints (Trujillo & Brown 2001; Petit et al. 2011; Gladman & Volk 2021; Fraser et al. 2024). We proceeded by analysing the orbital distributions of the classical objects obtained by all model systems. Confirming prior results in the literature (Section 1), we found that only the fiducial model yielded a concentration of classical objects at around $a \sim 44$–$45$ au; this is akin to the kernel (Figures 8 and 9). This conclusion held for the tested configurations of the outer-disk edge at 45 au and 47 au. Therefore, these results imply that a jumping Neptune explains kernel formation, even if the disk outer edge was at 45 au. This implies that the alternative hypothesis, that kernel formation involved only the outer edge of the disk, is not supported by our results.

Further investigations of the classical region, which require detailed analysis of the cold and hot populations and comparisons using a dedicated survey bias simulator, are beyond the scope of this paper. We are satisfied, however, that our models reproduce the bulk of the classical populations, including the kernel.

Moving beyond kernel formation in the classical region, we next investigated the effects of a jumping Neptune on other regions of the Kuiper Belt. Specifically, a comparison of the fiducial and 4GP models revealed that the former produced a smaller fraction of resonant objects in the trans-Neptunian region (roughly a 25% reduction), allowing resonant dropouts to contribute to both the scattered and detached populations (Tables 3 and A1). Therefore, although Neptune generally possessed resonant populations after the jump (e.g. via instantaneous capture of objects immediately following the jump and further captures during the subsequent outward migration to its present orbit), the net effect of this discontinuity was diminished occupancy of the 2:1 MMR and other resonances by the final populations.

To end this subsection, the remarkable similarities between our results and those of the literature imply that our fiducial model effectively captures the jumping Neptune scenario (e.g. Nesvorny 2015b).

### 3.4.3. The influence of primordial scattered disks

Models 4GP+SD and 4GP+hotSD included primordial scattered disks extending to 80 au.

These components enabled the 5:2, 3:1, and other distant MMRs to sweep and capture larger fractions of disk objects (Figures 5, 6, and 9). For example, in the 4GP+hotSD model, the 3:2-to-5:2 resonant population ratio was 3.0, the closest of all models to the observed datum. However, this value remains higher than the real intrinsic ratio. Reconciling this with observations may require a more populous, primordial scattered disk than that modelled here. However, this may create final systems with excessively large scattered populations. Another approach to the problem is to assess perturbations from primordial Plutos. These would deplete the 3:2 resonant population more significantly (see Section 3.4.1). Our results (Table 3) indicate that this would reduce the 3:2-to-5:2 ratio by a factor of 7.4/32, or ~ 0.23. As such perturbations would not suffice to align the 3:2-to-5:2 ratio with the observations of the 4GP and fiducial models, we conclude that only the 4GP+SD and 4GP+hotSD models may yield a 5:2 resonant population close to observed values. In other words, combining the effects of self-gravitating Plutos with a primordial scattered disk yields 3:2-to-5:2 ratios of 0.23 × 8.6 ~ 2 and 0.23 × 3.0 ~ 0.7, respectively, within the observed intrinsic range.

However, the 4GP+SD and 4GP+hotSD models have unique issues. As a significant fraction of primordial scattered disk objects survived after 4.5 Gyr (5.6% for 4GP+SD and 9.0% for 4GP+hotSD), these models produced overly large scattered populations compared to detached populations (Table 3). Moreover, the primordial scattered disk survivors influenced the inclination distribution beyond 50 au, such that the final fractions of high-$i$ objects were inconsistent with the observed intrinsic estimates (Figures 9 and 10; Table 3). Thus, to understand the orbital conditions of the protoplanetary disk better, it is essential to vary the population sizes and orbital properties of the primordial scattered objects.

3.4.4. The distant Kuiper belt: scattered, detached, and extreme populations

Turning to the orbital structure in the distant Kuiper Belt, our results confirm the key findings of standard Kuiper Belt formation models (Lykawka & Mukai 2007c; Gomes et al. 2008; Sheppard et al. 2016; Nesvorny & Vokrouhlicky 2016; Nesvorny et al. 2016; Kaib & Sheppard 2016; Pike et al. 2017; Kaib et al. 2019; the 'control model' of Lykawka & Ito 2023). First, we identified a population of stable, distant resonant objects comprising ~9.5% and ~14% of the populations in the $a > 50$ au region for models without and with primordial scattered disks, respectively. We also found some resonant dropouts near these resonances, which our algorithm classified as scattered or detached objects. Conversely, our results indicate that ~86–90.5% of the distant Kuiper Belt region consists of scattered and detached populations (including transient resonant objects). Therefore, the orbital properties of the scattered and detached populations impose crucial constraints on the distant trans-Neptunian region.

Can the models explored in this study explain the detached, high-$i$, and extreme TNO populations? In the analysis below, we emphasize the findings of the fiducial and 4GP_pluSG models that yielded the most promising results (as discussed in previous sections), but also mention other models for completeness. After investigating the orbital properties of the distant populations, we found that although scattered objects were concentrated well below the 40 au perihelion threshold, detached objects were concentrated at $q \sim$ 40–60 au. The detached population was confined within $a \sim$ 250–300 au. This is the region where n:1 and n:2 MMRs influence the long-term dynamics of scattered disk objects (Lykawka & Mukai 2006; Gallardo 2006; Graham & Volk 2024). Our results also reveal a remarkable correlation between perihelion and inclination, especially for objects of approximate $q >$ 37 au (Figure 11). Such an outcome was expected, given the vZLK-coupled MMR dynamics (Section 1). Furthermore, although scattered objects exhibited a broad range of inclinations, their detached counterparts were strongly concentrated at inclinations > 20°. Specifically, in line with the $q$–$i$ correlation discussed above, all obtained detached objects with $q >$ 40 and $q >$ 50–60 au were confined to $i >$ 20° and $i >$ 20–30° respectively. These results are consistent with previous studies on the dynamics of distant populations (Lykawka & Mukai 2007c; Gomes et al. 2008; Nesvorny et al. 2016; Lykawka & Ito 2023; Kaib et al. 2025). Moreover, the fiducial model yielded a scattered-to-detached population ratio of 3.7 in the distant Kuiper Belt. Although the model incorporating mutual gravitational effects among the Plutos yielded an improved ratio of ~2.9–3.0, this remains at least three times higher than the observationally constrained value (Table 3). In addition, a comparison with the orbital distribution of known distant TNOs (Figures 5, 6, 11, and 12) revealed that our models fail to explain the following observed subpopulations: detached TNOs with $a >$ 245 au, scattered TNOs with $q >$ 37 au and $i <$ 20°, detached TNOs with $i <$ 20°, and all known extreme TNOs with $q >$ 50 au. In conclusion, the results indicate that the fiducial and other standard models cannot account for the detached TNO population. Recently, Kaib et al. (2025) also concluded that even models incorporating additional perturbations from temporary, primordial Mars- to Earth-mass embryos fail to reproduce the observed, low-inclination detached population, including the extreme members thereof with $q >$ 50 au.

Turning to the highly inclined portion of the inclination distribution, the fiducial model yielded a high-$i$ fraction of 2.4% in the distant Kuiper Belt (Table 3). The 4GP and 4GP_pluSG models exhibited slightly higher fractions of 3.3% and 2.9%. However, the models incorporating a primordial scattered disk yielded excessively low fractions (0.9% and 0.2%). Recall that the observed 2% fraction of high-$i$ TNOs is apparent (Section 1). Therefore, even our best estimates of 2.4–3.3% are likely too low, as the intrinsic (unbiased) fraction must substantially exceed 2%, given the observational biases against detection of higher-inclination TNOs. Separate analyses of the high-$i$ fractions in the scattered and detached populations yielded similar findings. These results indicate that the fiducial and other standard models do not generate the highly inclined populations that are required to explain the observed high-$i$ TNOs.

Turning to extreme TNOs (Table 1), we found that our models produced some TNOs with $q > 60$ au or $i \sim 50$–$60°$ ($q < 40$ au), albeit with low efficiencies of ~0.16% and ~0.06%. These fractions are a few times smaller than the apparent observed values, 0.5% (4/804) and 0.2% (8/3725), respectively[14]. Recalling that observational biases hinder the discovery of high-$i$ or large-$q$ TNOs (Gladman et al. 2012; Lawler et al. 2017; Shankman et al. 2017; Kavelaars et al. 2020; Gladman & Volk 2021), the intrinsic fractions of these extreme TNO populations must be substantially higher. Another issue is that the extreme objects with $q > 50$ au that formed within our simulations were concentrated in the parameter space $a \sim 50$–$200$ au and $i = 30$–$55°$. However, the six currently known extreme TNOs with $q > 50$ au do not occupy that particular region of orbital space. Additionally, seven extreme TNOs on orbits with $a > 245$ au are known. Furthermore, our models proved highly inefficient at producing extreme objects with $q > 50$ au and $i < 20°$, or those with $i > 60°$, yielding zero or near-zero representatives in both categories among the ~100000 surviving objects after combining all final systems. These results demonstrate that our models failed to account for the existence of a significant population of extreme TNOs.

In conclusion, standard scenarios that consider only the four giant planets, as tested in our five models, including the state-of-the-art fiducial model with a jumping Neptune, failed to reproduce the detached and high-$i$ populations, as well as their extreme members. Therefore, additional mechanisms (past or ongoing) in the solar system are required to explain these populations consistently. Proposed hypotheses include perturbations from rogue planets (Huang et al. 2022b; but see (*in contra*) Kaib et al. 2025), undiscovered distant planets (Lykawka & Mukai 2008; Batygin & Brown 2016), and galactic tides combined with passing stars in the early solar system (Nesvorny et al. 2023). The findings of Lykawka & Ito (2023), who conducted a comprehensive analysis, imply that the gravitational perturbations imparted by an undiscovered Earth-class planet would offer a plausible and promising explanation.

### 4. Summary

Using suites of N-body simulations, our baseline model investigated the 4.5 Gyr orbital evolution of early systems comprising a protoplanetary disk, the four giant planets, and 1500 primordial Pluto-class bodies ("Plutos") that triggered the grainy migration of Neptune. We also considered models incorporating the jumping Neptune scenario, primordial scattered disks, or self-gravitating Plutos. The primary goal was to evaluate whether these models replicated the distant Kuiper Belt's orbital architecture at $a > 50$ au while remaining consistent with known constraints within 50 au.

---

[14] We considered all TNOs for high-$i$ fractions and distant TNOs ($a > 50$ au) for population fractions with $q > 60$ au.

Overall, our models produced systems that contained the main dynamical populations of the Kuiper Belt: cold/hot classical, resonant, scattered, and detached. From 67 final systems across all models, we identified a substantial Gyr-stable resonant population concentrated in close Neptunian MMRs (e.g. 1:1, 3:2, 5:3, 7:4, and 2:1) but with notable declines in occupancy beyond ~50 au (e.g. 5:2 and more distant MMRs). The best results were obtained when Neptune jumped in the early solar system, primordial self-gravitating Plutos induced grainy migration of Neptune, and the outer edge of the primordial disk lay at ~45–47 au. This specific scenario could replicate the kernel structure in the classical region and produce a less prominent resonant population that was more aligned with observations. However, the strength and duration of graininess depended on the initial abundance and decay of Plutos. The effect was primarily early and diminished as Plutos were removed. Notably, our results imply that the initial number of primordial Plutos should be ~150–500, implying that future investigations of Neptune's grainy migration should account for the gravitational effects of a few hundred such objects to confirm these findings. Finally, models incorporating a primordial scattered disk produced higher populations in distant MMRs (e.g. the 5:2 and n:1 MMRs). However, these models yielded overpopulations of scattered disk objects, inconsistent with observations.

No model efficiently generated detached objects with $q > 40$ au or high-$i$ objects with $i > 45°$. The population fractions were well below the expected (observed) intrinsic values. Additionally, the models failed to reproduce several known subpopulations of orbital space: scattered TNOs with $q > 37$ au and $i < 20°$, distant detached TNOs with $a > 245$ au, and detached TNOs with $i < 20°$. Furthermore, no model produced a population of extreme objects well, defined by $q > 50$ au or $i > 50°$.

In summary, although standard four-giant-planet scenarios such as those considered in this study potentially reproduce the known state of the trans-Neptunian region to 50 au, they fail to explain the populations and orbital properties of detached and high-$i$ TNOs, especially the extreme members thereof. This implies that additional mechanisms, such as perturbations from an unseen planet or planets, or early passing stars, may be needed to explain such populations. Future models must incorporate these mechanisms and a realistic grainy migration driven by hundreds of self-gravitating Plutos to elucidate the orbital structure of the outer solar system better. Such efforts are timely, coinciding, as they do, with the late-2025 launch of LSST operations (Ivezic et al. 2019; Kurlander et al. 2025). This 10-year survey will map high ecliptic latitudes, where high-$i$ TNOs are more detectable than elsewhere, despite their lower density relative to ecliptic surveys.

## Acknowledgments

We are grateful to reviewer Nathan A. Kaib for insightful comments that greatly enhanced the clarity and presentation of this work. The simulations presented here were partially performed using the general-purpose PC cluster at the Center for Computational Astrophysics in the National Astronomical Observatory of Japan. We are grateful for the generous time allocated to run the simulations. This work was supported by JSPS KAKENHI Grant Number JP23K03482. P.H.B. acknowledges support from the DIRAC Institute in the Department of Astronomy at the University of Washington. The DIRAC Institute is supported through generous gifts from the Charles and Lisa Simonyi Fund for Arts and Sciences, and the Washington Research Foundation.

## Data Availability

The presented data substantiate this study's findings. Observational data is available at https://newton.spacedys.com/astdys/. Supplementary data are available upon reasonable request from the corresponding author.

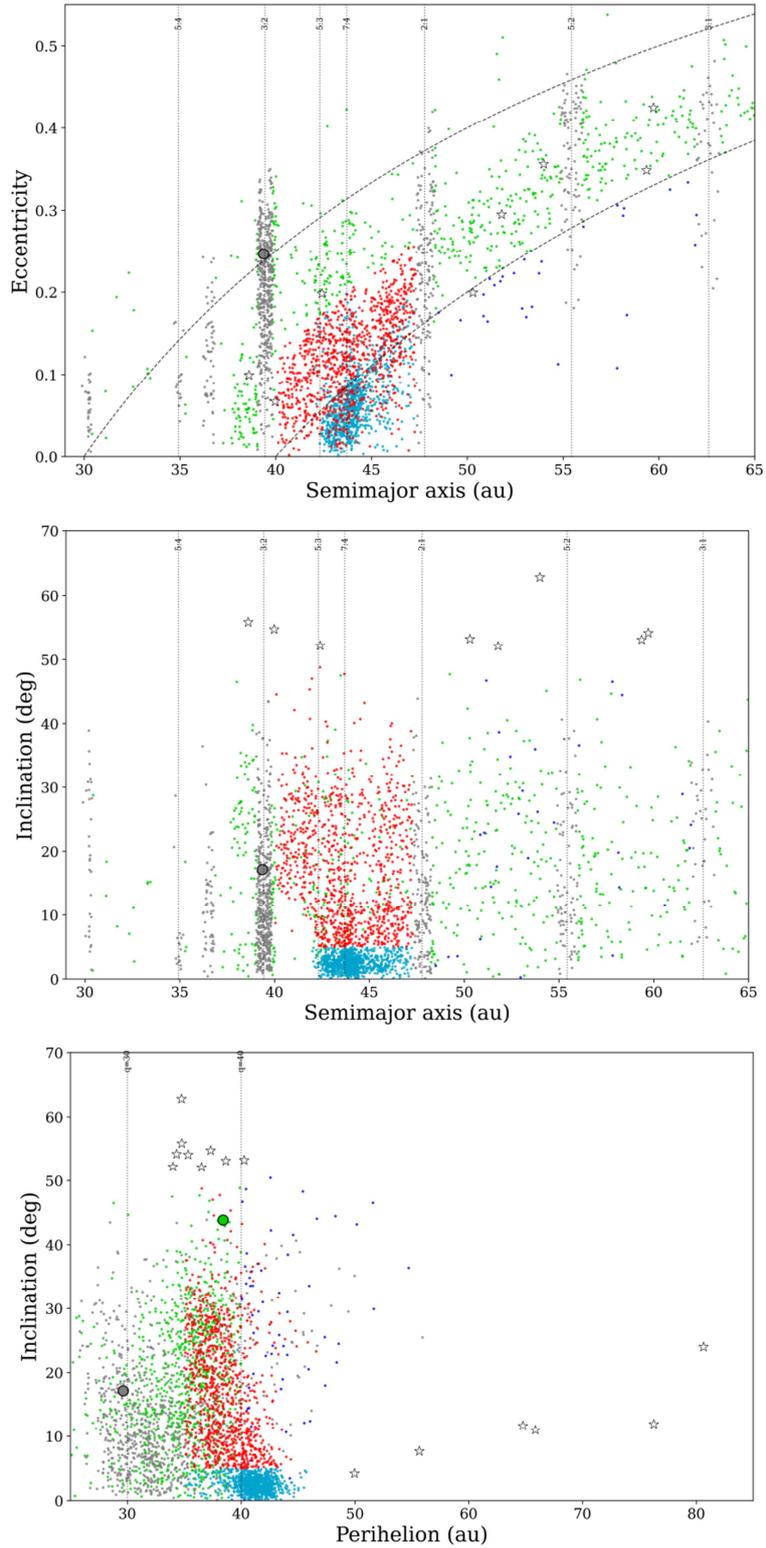

**Figure 1.** Orbital distribution of currently known trans-Neptunian objects (TNOs) with small orbital uncertainties. TNOs are dynamically classified as potentially resonant (gray), cold classical (dark cyan), hot classical (red), scattered (green), and detached (blue). Extreme TNOs are indicated by stars (Table 1). Pluto and Eris are shown using big circles and color-coded according to the same classification. Curves indicate Neptunian mean motion resonances and perihelion distances of 30 au and 40 au. See Sections 1 and 2 for more details.

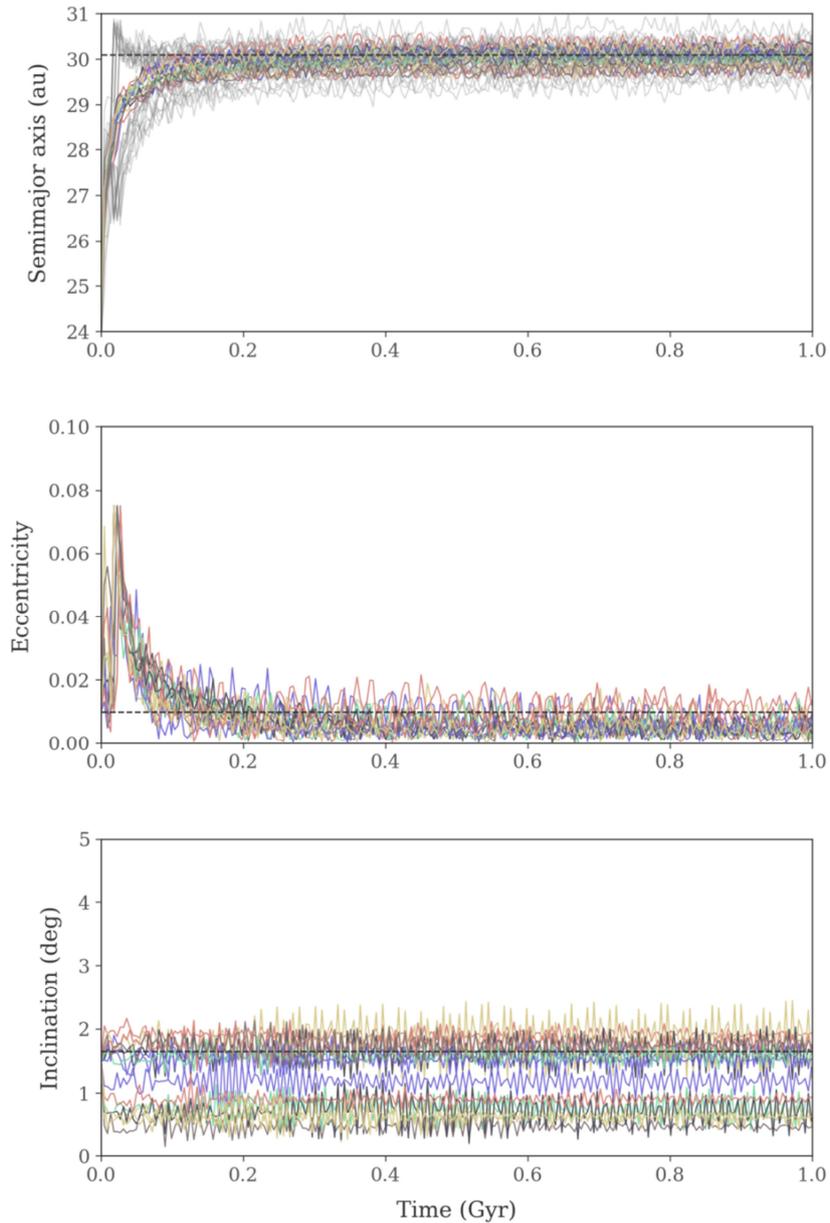

**Figure 2.** Orbital evolution of 15 Neptune analogs over the first 1 Gyr in the fiducial model (grainy migration plus a jumping Neptune). All analogs exhibit outward migration from 24 au to ~30 au, accompanied by eccentricity damping from initial values 0.075 to nearly circular orbits, and inclination evolution toward low-inclination configurations. Each colored line represents one Neptune analog. The gray lines in the top panel represent the evolution of perihelia and aphelia of the 15 Neptune analogs. The dashed lines indicate the mean values for Neptune over the last 100 Myr.

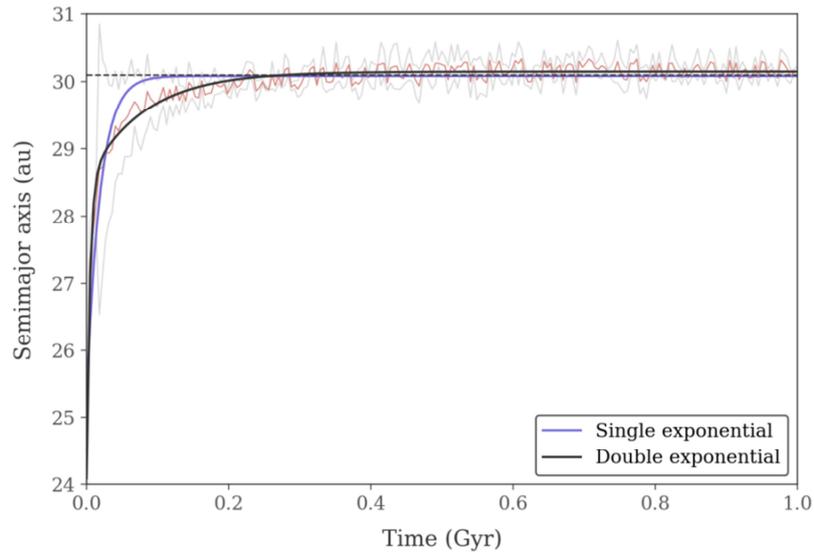

**Figure 3.** Orbital evolution of a Neptune analog from the fiducial model over the first 1 Gyr (red). Two exponential decay models are compared: single exponential (purple) and double exponential (black). In this example, the double exponential model more accurately captures the two-phase evolution, characterized by a rapid initial decay (tau1 = 5 Myr) followed by gradual stabilization (tau2 = 80 Myr) toward its final orbit. The gray lines in the top panel represent the evolution of perihelia and aphelia of the Neptune analog. The dashed line indicates the mean semimajor axis for Neptune over the last 100 Myr.

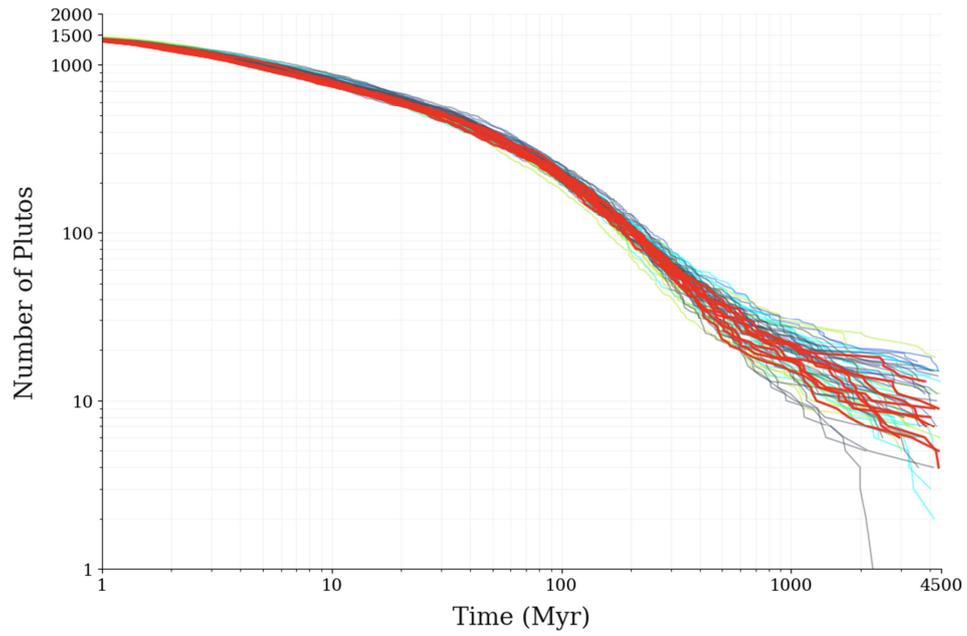

**Figure 4.** Decay of Pluto-class objects over time for all systems modeled in this work. Each curve shows the decay for a single system within a specific model (Table 2). Colors denote distinct models (4GP in dark gray; 4G_pluSG in red; 4GP_jNep in cyan; 4GP+SD in blue; 4GP+hotSD in lime).

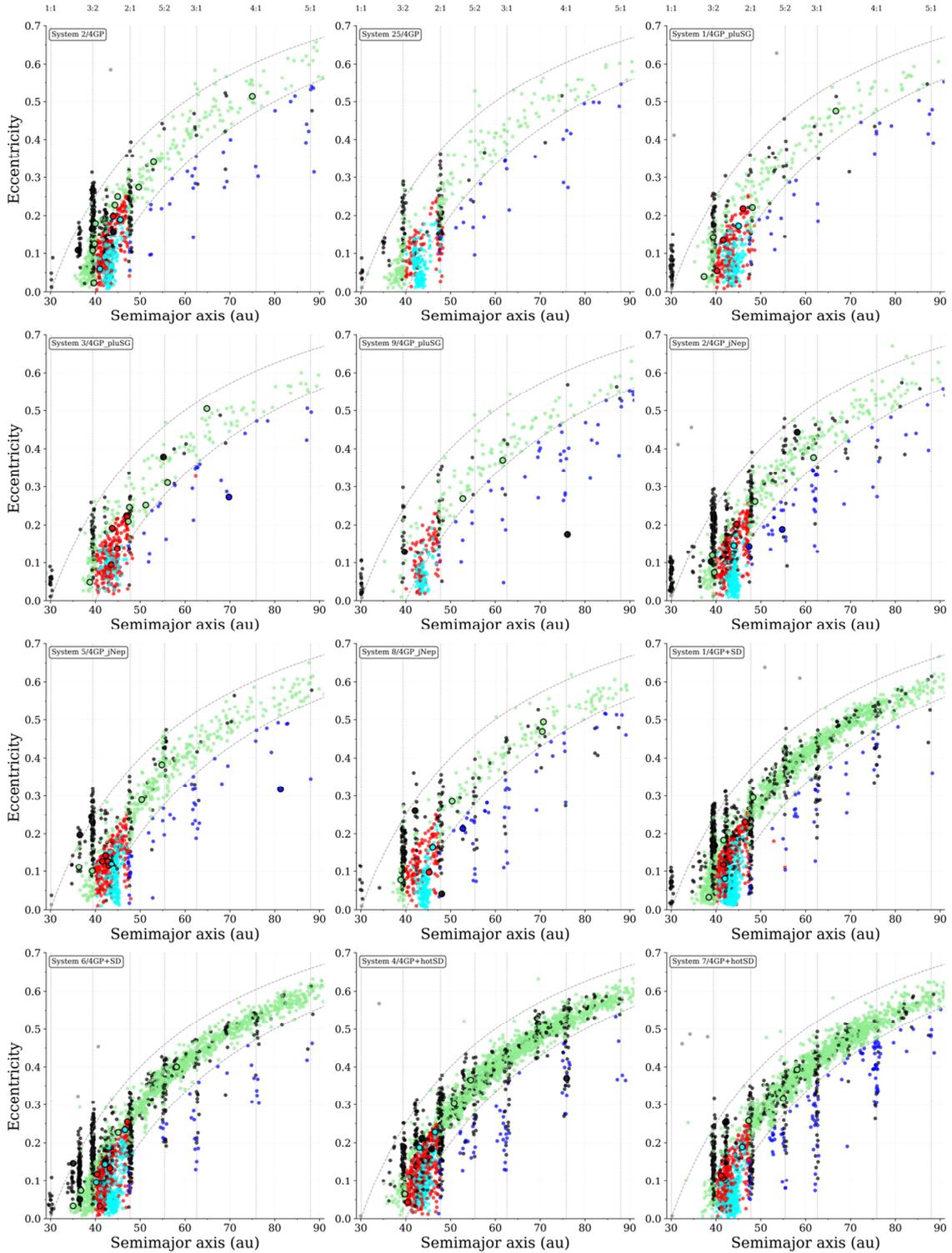

**Figure 5.** Orbital distribution of objects surviving 4.5 Gyr of dynamical evolution across representative systems obtained in our planetary migration models (Table 2). Each panel shows survivors from individual systems. Objects are color-coded by dynamical class: resonant (black), cold classical (cyan), hot classical (red), scattered (light green), and detached (blue). A few survivors are indicated as Centaurs (gray) and detached objects with $i < 5°$ and $a < 70$ au (dark yellow). Pluto-class objects are shown with larger symbols. Curves indicate Neptunian mean motion resonances and perihelion distances of 30 au and 40 au.

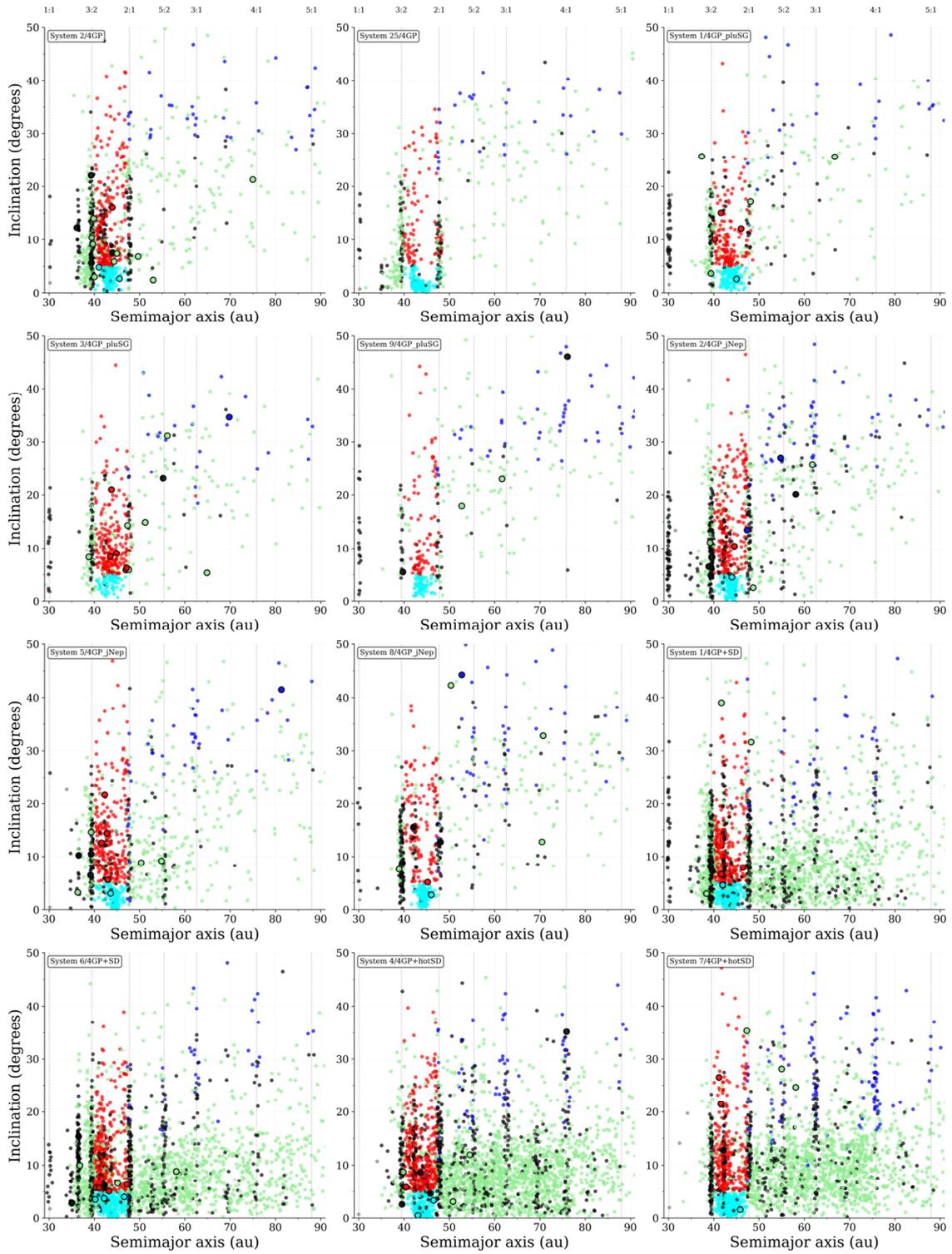

**Figure 6.** Similar to Figure 5, but showing the inclinations of the objects.

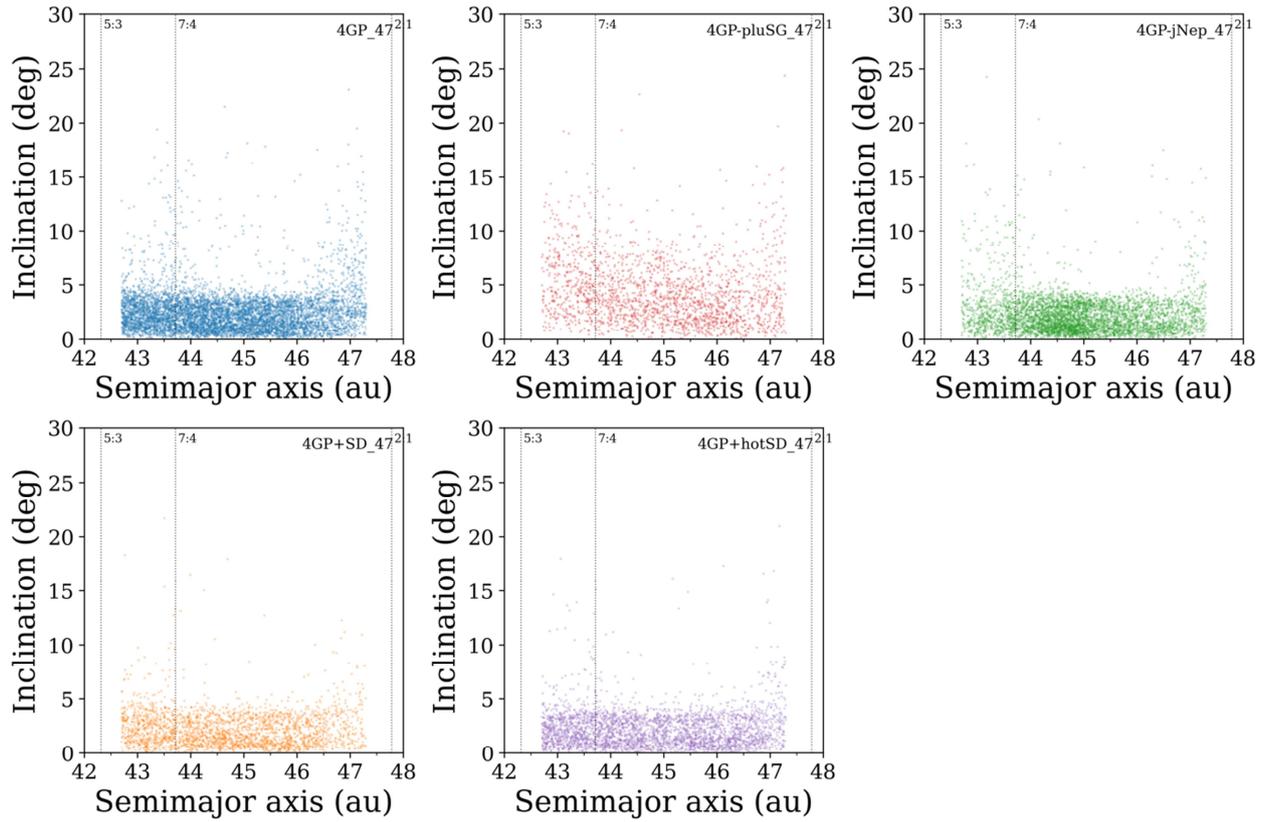

**Figure 7.** Inclination distribution of local (outer disk) objects that acquired cold classical orbits within the range $a = 42.7–47.3$ au, as obtained from five planetary migration models (Table 2). Each panel shows results combined from the individual final systems of the 4GP (26 systems), 4GP_pluSG (9 systems), fiducial 4GP_jNep (15 systems), 4GP+SD (8 systems), and 4GP+hotSD (9 systems) models. Vertical dotted lines mark Neptunian mean motion resonances.

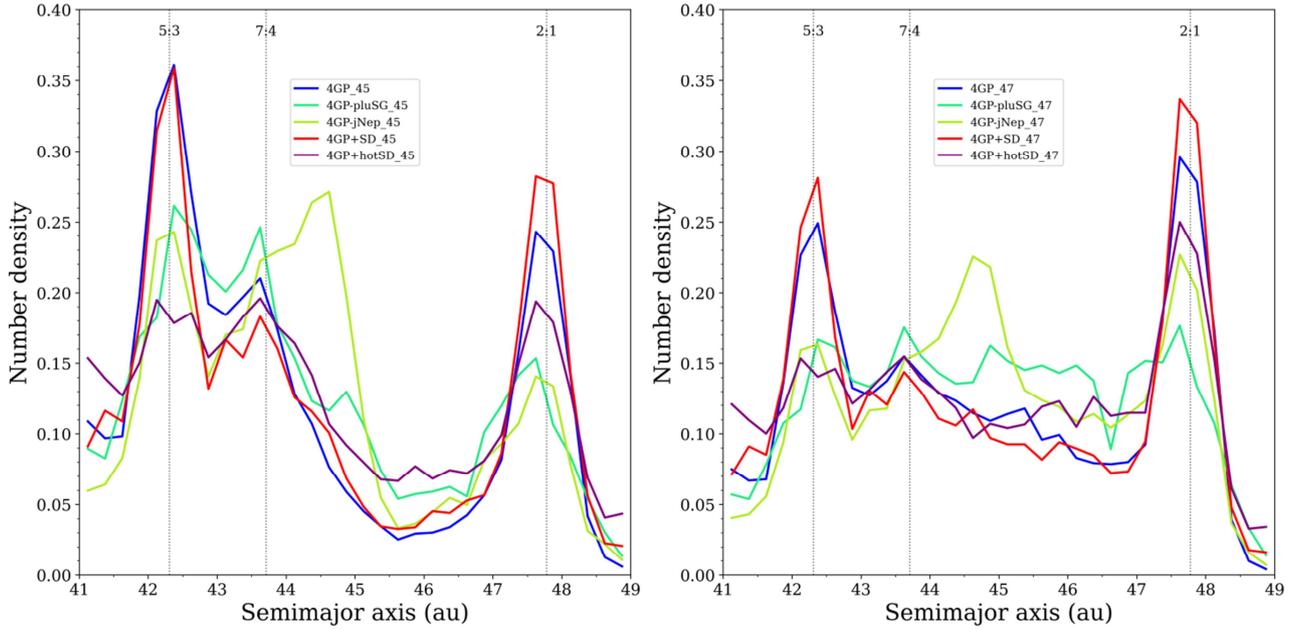

**Figure 8.** Normalized number density distributions of classical objects for protoplanetary outer disks extended to 45 au (left panel) and 47 au (right panel). The distributions cover five models after combining their final systems: 4GP (26 systems) (blue), 4GP_pluSG (9 systems) (mint green), fiducial 4GP_jNep (15 systems) (lime green), 4GP+SD (8 systems) (red), and 4GP+hotSD (9 systems) (purple). Densities are calculated using 0.25 au bins and normalized within each disk extension case for comparison. Vertical dotted lines mark major mean motion resonances with Neptune. The kernel concentration of classical objects is visible only for the fiducial model at $a \sim$ 44–45 au in both panels.

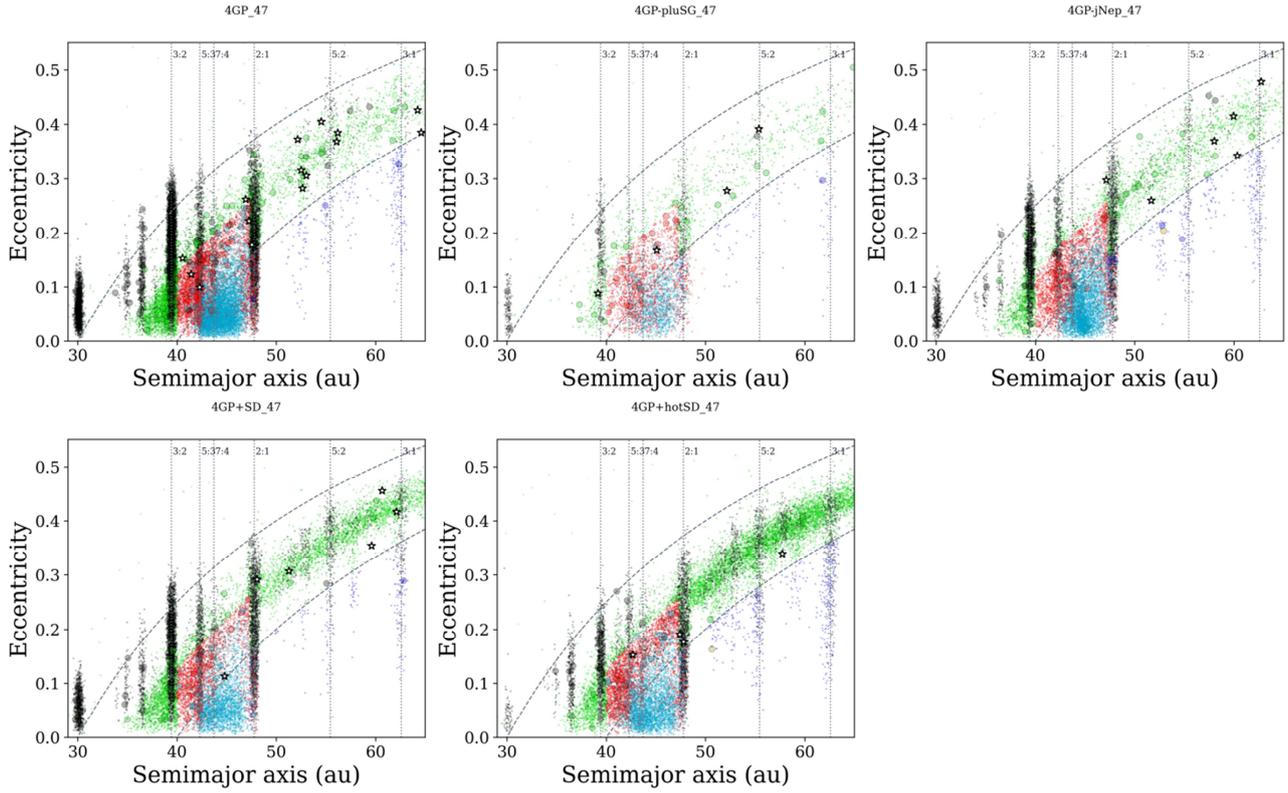

**Figure 9.** Orbital distribution of objects across five planetary migration models (Table 2). We combined the final systems in each model presented in this figure: 4GP (26 systems), 4GP_pluSG (9 systems), fiducial 4GP_jNep (15 systems), 4GP+SD (8 systems), and 4GP+hotSD (9 systems). White stars indicate extreme objects that meet the criteria for high perihelion and/or inclination (Section 1). Color-coded dynamical classification and the curves are the same as in Figure 5.

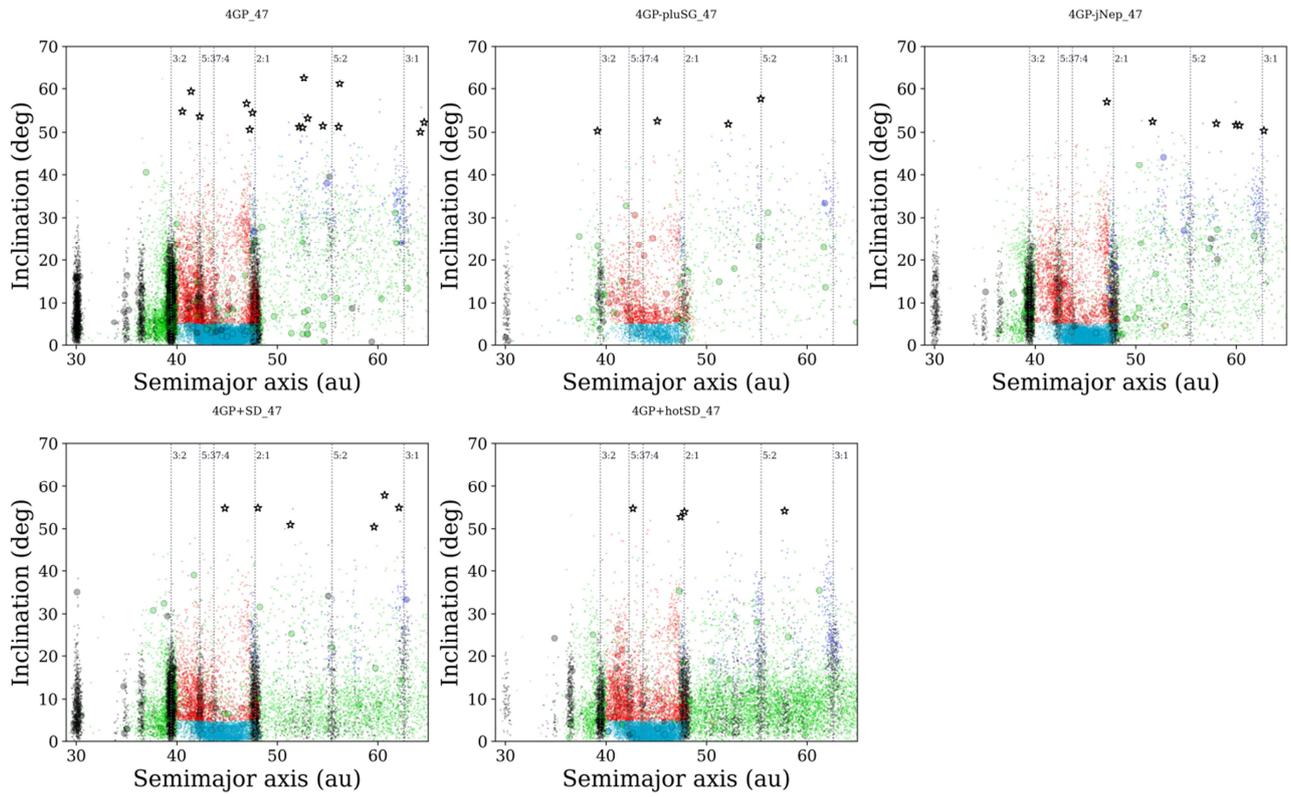

**Figure 10.** Similar to Figure 9, but showing the inclinations of the objects.

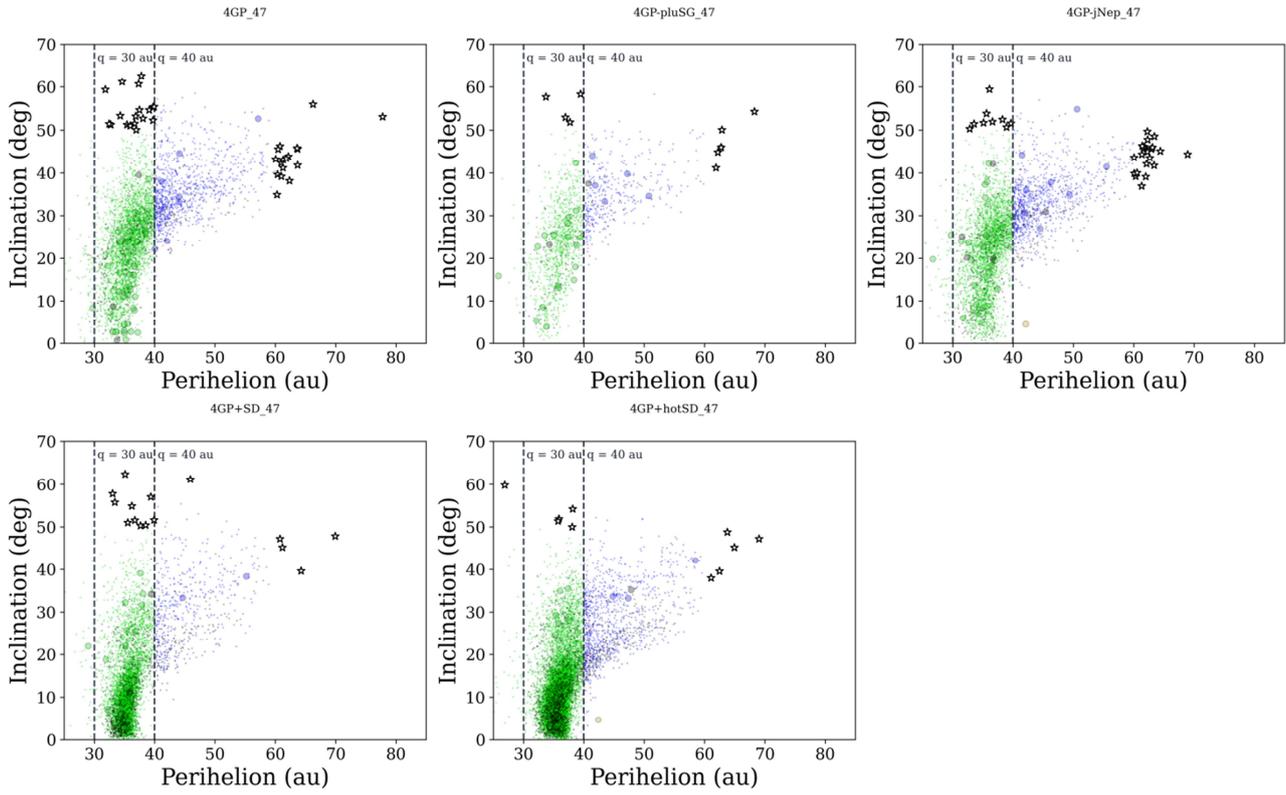

**Figure 11.** The same as in Figure 9, but showing the perihelion vs. inclination distribution of distant objects that survived with $a > 50$ au.

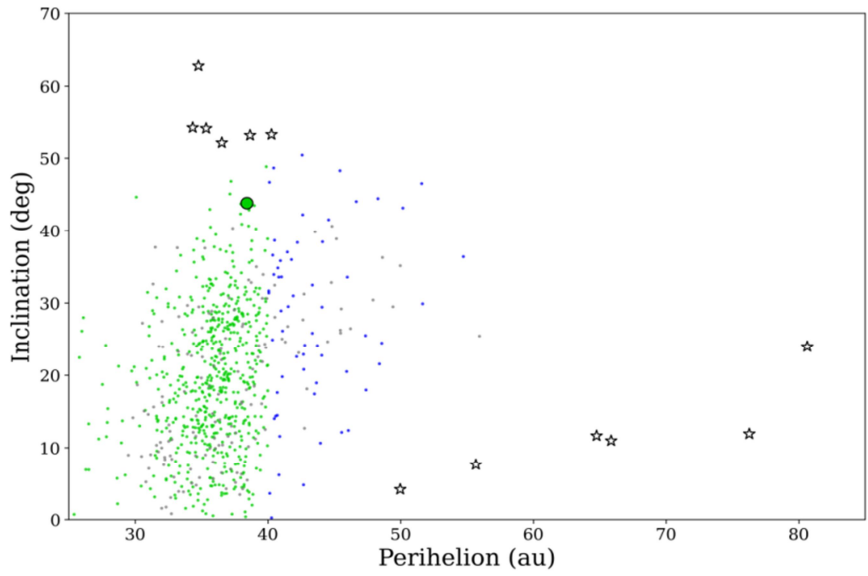

**Figure 12.** The same as in Figure 1, but showing only distant TNOs with *a* > 50 au in perihelion vs. inclination orbital space.

**Table 1.** Extreme trans-Neptunian objects (TNOs).

| Object | $a$ (au) | $i$ (°) | $q$ (au) | Parent population | Criterion | Object reference |
|---|---|---|---|---|---|---|
| 2023 KQ$_{14}$ | 245.51 | 11.0 | 65.87 | detached | $q \gtrsim 60$ au | 1 |
| 2012 VP$_{113}$ | 271.39 | 24.0 | 80.62 | detached | $q \gtrsim 60$ au | 2 |
| (90377) Sedna | 551.96 | 11.9 | 76.26 | detached | $q \gtrsim 60$ au | 3 |
| (541132) Leleakuhonua | 1451.44 | 11.7 | 64.76 | detached | $q \gtrsim 60$ au | 4 |
| 2013 SY$_{99}$ | 850.12 | 4.2 | 49.96 | detached | $q \gtrsim 50$ au, $i \lesssim 20°$ | 5 |
| 2021 RR$_{205}$ | 1012.26 | 7.6 | 55.66 | detached | $q \gtrsim 50$ au, $i \lesssim 20°$ | 6 |
| 2014 YX$_{91}$ | 53.99 | 62.8 | 34.77 | high-$i$ | $i \gtrsim 60°$ | 7 |
| 2016 RJ$_{82}$ | 38.61 | 55.9 | 34.78 | high-$i$ | $i \gtrsim 50°$, $q \lesssim 40$ au | - |
| 2014 YK$_{92}$ | 39.97 | 54.8 | 37.31 | high-$i$ | $i \gtrsim 50°$, $q \lesssim 40$ au | - |
| 2015 XO$_{480}$ | 42.41 | 52.2 | 34.01 | high-$i$ | $i \gtrsim 50°$, $q \lesssim 40$ au | - |
| 2017 EJ$_{52}$ | 50.30 | 53.3 | 40.27 | high-$i$ | $i \gtrsim 50°$, $q \lesssim 40$ au | 8 |
| 2015 UO$_{105}$ | 51.79 | 52.1 | 36.51 | high-$i$ | $i \gtrsim 50°$, $q \lesssim 40$ au | - |
| 2014 UN$_{225}$ | 59.35 | 53.2 | 38.66 | high-$i$ | $i \gtrsim 50°$, $q \lesssim 40$ au | 9 |
| 2017 FO$_{161}$ | 59.71 | 54.2 | 34.33 | high-$i$ | $i \gtrsim 50°$, $q \lesssim 40$ au | 6 |
| (768325) 2015 BP$_{519}$ | 493.43 | 54.1 | 35.34 | high-$i$ | $i \gtrsim 50°$, $q \lesssim 40$ au | 10 |

**Notes.** $a$, $i$, and $q$ represent the TNO's semimajor axis, inclination, and perihelion, respectively. We considered that a TNO must have $a > 30$ au and $q > 25$ au. The origin of extreme TNOs with $q > 60$ au or $i > 60°$ likely requires additional perturbations other than the four known solar system's giant planets. The orbits of all extreme TNOs are unlikely to originate via resonant interactions with the giant planets. See Section 1 for more details. Orbital data retrieved from the AstDys observational database at MJD = 60800 in July 2025. References: (1) Chen et al. (2025); (2) Trujillo & Sheppard (2014); (3) Brown et al. (2004); (4) Sheppard et al. (2019); (5) Bannister et al. (2017); (6) Sheppard et al. (2022); (7) Bernardinelli et al. (2022); (8) Deen (2024); (9) Elliott et al. (2017); (10) Becker et al. (2018).

**Table 2.** Summary of simulation initial conditions of the models considered in this work.

| | Model | Inner disk | | Outer disk | #Runs |
|---|---|---|---|---|---|
| ID | Short description | Massive component | Massless component | | |
| 4GP | Four giant planets, Plutos | 1500P+14250 [14.5] | 136000 (86000) | 2500 (1800) | 20 (10) |
| 4GP_pluSG | Four giant planets, self-gravitating Plutos | 1500P+14250 [14.5] | 86000 | 1800 | 10 |
| 4GP_jNep (fiducial) | Four giant planets, Plutos, jumping Neptune | 1500P+11250 [12.5] | 139000 | 2500 | 15 |
| 4GP+SD | Four giant planets, Plutos, primordial scattered disk | 1500P+14250 [14.5] | 141000 | 2500 | 10 |
| 4GP+hotSD | Four giant planets, Plutos, hot primordial scattered disk | 1500P+15000 [15.0] | 140000 | 2500 | 10 |

**Notes.** All models began with the four giant planets — Jupiter, Saturn, Uranus, and Neptune—in a spatially compact configuration. The protoplanetary disk was represented by two main components: the inner and outer disks, initially set with orbits at $a = 25$–$30$ au ($e < 0.2$ and $i < 5°$) and $a = 30$–$49$ au ($e < 0.01$ and $i < 0.5°$), respectively. The inner and outer disk objects were distributed following a number density proportional to $a^0$ (uniform) and $a^{-1}$, respectively. In addition to the massive inner disk objects, 1500 Pluto-class objects on dynamically cold orbits ($e < 0.01$ and $i < 0.5°$) were placed uniformly across 25–30 au, totaling 5 ME of the massive inner disk component. After interacting with the massive disk, the giant planets migrated to their current orbits and evolved over a period of 4.5 Gyr. The number within brackets represents the total mass of the massive component, while the numbers within parentheses indicate additional low-resolution runs. In the 4GP_pluSG model, the 1500 Plutos gravitationally interacted with each other and all other bodies in the system. In the 4GP_jNep model, Neptune jumped during planet migration to mimic the giant-planet instability. The 4GP+SD and 4GP+hotSD models contained 15% of the massless component placed at $a = 30$–$80$ au and $q = 24$–$35$ au (with inclinations the same as for the inner disk objects) to simulate a primordial scattered disk. In the 4GP+hotSD model, all the inner disk objects started with $e < 0.35$ and $i < 10°$. See Section 2 for other details.

**Table 3.** Summary of key properties of the final systems obtained after 4.5 Gyr for each model considered in this work. The outer region of the protoplanetary disk was assumed to extend up to 45 au.

| | Model | | | | | Observations | |
|---|---|---|---|---|---|---|---|
| | 4GP | 4GP_pluSG | 4GP_jNep | 4GP+SD | 4GP+hotSD | Intrinsic | Reference |
| Plutos | 8.5 ± 5.8 | 7 ± 3 | 8 ± 5 | 15 ± 4 | 7 ± 5 | ≥2 | |
| Plutos ($a$ = 30–50 au) | 7.5 ± 4.8 | 4 ± 3 | 5 ± 4.5 | 11 ± 3 | 6 ± 3 | 1 | |
| Plutos ($a$ = 30–100 au) | 8 ± 5.5 | 6 ± 4 | 7 ± 7 | 14 ± 4.3 | 7 ± 5 | 2 | |
| $r_{32,21}$ | 2.9 ± 1.5 | 1.6 ± 2 | 3.7 ± 1.9 | 2.1 ± 1 | 1.1 ± 0.4 | 0.7-4 | 1,2,3,4,5,6 |
| $r_{HC,32}$ | 0.7 ± 0.4 | 5.1 ± 3.4 | 1.1 ± 0.5 | 0.8 ± 0.3 | 2.5 ± 0.9 | 2-4.7 | 1,2,3,4,6,7,8,9 |
| $r_{32,52}$ | 32.0 ± 33.8 | 7.4 ± 20.8 | 16.8 ± 14.4 | 8.6 ± 7.1 | 3.0 ± 3.4 | 0.9-2 | 1,2,3,4,5,6 |
| $\sigma_{32}$ (°) | 7.8 | 9.5 | 8.5 | 7.8 | 6.8 | 11-16 | 1,10 |
| $\sigma_{21}$ (°) | 8.2 | 9.7 | 10.3 | 8.4 | 8.4 | 6-7 | 1,11 |
| $\sigma_{52}$ (°) | 16.8 | 18.2 | 15.0 | 9.8 | 10.6 | 14 | 1 |
| $f$-Centaur (%) | 0.3 ± 0.2 | 0.6 ± 0.5 | 0.2 ± 0.1 | 0.3 ± 0.6 | 0.2 ± 0.1 | — | |
| $f$-resonant (%) | 38.0 ± 13.6 | 13.8 ± 5.9 | 28.2 ± 7.6 | 35.4 ± 11.5 | 20.0 ± 4.1 | ~13-20 | 6,12 |
| $f$-classical (%) | 31.0 ± 13.4 | 42.7 ± 10.6 | 31.7 ± 2.9 | 20.5 ± 3.7 | 21.4 ± 2.6 | ~17-22 | 6,12 |
| $f$-classical cold (%) | 18.0 ± 8.2 | 13.8 ± 5.2 | 18.0 ± 2.4 | 11.0 ± 1.9 | 11.8 ± 2.3 | — | |
| $f$-classical hot (%) | 11.9 ± 4.2 | 23.5 ± 9.5 | 13.6 ± 1.6 | 9.6 ± 1 | 9.8 ± 0.5 | — | |
| $f$-scattered (%) | 28.6 ± 3.2 | 35.5 ± 6.4 | 32.3 ± 2.3 | 41.2 ± 6.8 | 55.6 ± 5.7 | ~45-70* | 6,12 |
| $f$-detached (%) | 3.5 ± 1.6 | 6.9 ± 1.9 | 5.7 ± 2.5 | 2.4 ± 1.1 | 4.1 ± 1.9 | | |
| $r_{sca,det}$ | 3.9 ± 1 | 2.9 ± 0.5 | 3.7 ± 1.5 | 12.9 ± 4.7 | 11.0 ± 3.5 | ≤ 1.0 | |
| Fhi (%) | 3.3 ± 1.2 | 2.9 ± 1.7 | 2.4 ± 1.2 | 0.9 ± 0.6 | 0.2 ± 0.3 | ≥ 2.0 | |
| Fhi$_{sca}$ (%) | 1.4 ± 1.7 | 0.9 ± 0.9 | 0.8 ± 0.8 | 0.3 ± 0.4 | 0.1 ± 0.1 | ≥ 1.0 | |
| Fhi$_{det}$ (%) | 11.5 ± 7.8 | 10.3 ± 5.7 | 6.7 ± 6 | 5.3 ± 5.1 | 1.8 ± 2.1 | ≥ 7.0 | |
| #runs combined | 26 | 9 | 15 | 8 | 9 | | |

**Notes.** Plutos represent objects with masses comparable to that of Pluto. $r_{32,21}$, $r_{HC,32}$, and $r_{32,52}$ are the ratios of 3:2 to 2:1 resonant objects, hot classical to 3:2 resonant objects, and 3:2 to 5:2 resonant objects. $\sigma_{xy}$ is the inclination dispersion of a particular resonant population described by the x:y ratio (modelled assuming a PDF $\propto \sin(i) \times \exp(-i^2 / (2\sigma^2))$, $f$ represents the fraction of a particular population to the entire population of objects. $r_{sca,det}$ is the ratio of scattered and detached populations. Fhi represents the fraction of high-$i$ ($i > 45°$) objects. The subscripts 'sca' and 'det' refer to results for the scattering and

detached populations, respectively. The models are described in Table 2. Objects with $q > 25$ au were considered when evaluating the results reported in this table. The exceptions are $r_{sca,det}$, Fhi, Fhi$_{sca}$, and Fhi$_{det}$, where we obtained the results by adding the filter $a > 50$ au (distant objects only). Except for $\sigma_{xy}$, all results are shown using median value ± standard interquartile range IQR. The intrinsic value intervals reflect the distinct nominal results and limits found in the listed references. *This intrinsic interval consists of scattered and detached populations summed since distinct definitions are used in the literature. Intrinsic values and other related constraints are also discussed in Section 1 of this work. References: (1) Gladman et al. (2012); (2) Adams et al. (2014); (3) Nesvorny & Vokrouhlicky (2016); (4) Volk et al. (2016); (5) Pike et al. (2017); (6) Bernardinelli et al. (2025); (7) Volk & Malhotra (2019); (8) Petit et al. (2023); (9) Kaib et al. (2024); (10) Gulbis et al. (2010); (11) Chen et al. (2019); (12) Kurlander et al. (2025).

# Appendix

**Table A1.** Summary of key properties of the final systems obtained after 4.5 Gyr for the models considered in this work. The outer region of the protoplanetary disk was assumed to extend up to 47 au.

| | Model | | | | | Observations | |
|---|---|---|---|---|---|---|---|
| | 4GP | 4GP_pluSG | 4GP_jNep | 4GP+SD | 4GP+hotSD | Intrinsic | Reference |
| Plutos | 8.5 ± 5.8 | 7 ± 3 | 8 ± 5 | 15 ± 4 | 7 ± 5 | ≥2 | |
| Plutos ($a$ = 30–50 au) | 7.5 ± 4.8 | 4 ± 3 | 5 ± 4.5 | 11 ± 3 | 6 ± 3 | 1 | |
| Plutos ($a$ = 30–100 au) | 8 ± 5.5 | 6 ± 4 | 7 ± 7 | 14 ± 4.3 | 7 ± 5 | 2 | |
| $r_{32,21}$ | 1.8 ± 0.5 | 1.0 ± 0.6 | 1.6 ± 0.5 | 1.4 ± 0.4 | 0.7 ± 0.3 | 0.7-4 | 1,2,3,4,5,6 |
| $r_{HC,32}$ | 0.7 ± 0.4 | 6.3 ± 3.3 | 1.2 ± 0.6 | 0.8 ± 0.3 | 2.6 ± 0.7 | 2-4.7 | 1,2,3,4,6,7,8,9 |
| $r_{32,52}$ | 32.0 ± 33.8 | 7.4 ± 20.8 | 16.9 ± 14.4 | 8.6 ± 7.1 | 3.0 ± 3.4 | 0.9-2 | 1,2,3,4,5,6 |
| $\sigma_{32}$ (°) | 9.3 ± 1.7 | 12.3 ± 2.2 | 9.6 ± 1.1 | 9.2 ± 1.2 | 7.4 ± 2.1 | 11-16 | 1,10 |
| $\sigma_{21}$ (°) | 6.5 ± 2.2 | 9.4 ± 2.5 | 7.7 ± 2 | 7.7 ± 1.4 | 7.7 ± 1.3 | 6-7 | 1,11 |
| $\sigma_{52}$ (°) | 25.1 ± 8.4 | 19.6 ± 6.3 | 20.2 ± 4 | 8.5 ± 4.7 | 13.6 ± 5.6 | 14 | 1 |
| $f$-Centaur (%) | 0.2 ± 0.2 | 0.5 ± 0.6 | 0.2 ± 0.1 | 0.3 ± 0.5 | 0.2 ± 0.1 | — | |
| $f$-resonant (%) | 35.7 ± 11.8 | 11.9 ± 4.7 | 27.4 ± 7.6 | 35.1 ± 10.6 | 20.0 ± 4.8 | ~13-20 | 6,12 |
| $f$-classical (%) | 37.0 ± 11.7 | 50.8 ± 7 | 40.1 ± 3.5 | 25.0 ± 3 | 25.6 ± 3.9 | ~17-22 | 6,12 |
| $\quad f$-classical cold (%) | 24.9 ± 7 | 25.5 ± 7.7 | 27.9 ± 3.6 | 15.4 ± 2 | 16.3 ± 2 | — | |
| $\quad f$-classical hot (%) | 10.6 ± 3.8 | 22.2 ± 8.4 | 12.3 ± 1.3 | 9.3 ± 1.3 | 9.5 ± 0.4 | — | |
| $f$-scattered (%) | 24.2 ± 2.8 | 28.7 ± 3.1 | 27.4 ± 2 | 38.0 ± 7.2 | 51.6 ± 6.2 | ~45-70* | 6,12 |
| $f$-detached (%) | 3.0 ± 1.3 | 5.5 ± 1.6 | 4.6 ± 1.8 | 2.2 ± 0.9 | 3.8 ± 1.8 | | |
| $r_{sca,det}$ | 3.9 ± 1 | 3.0 ± 0.5 | 3.7 ± 1.5 | 12.9 ± 4.7 | 11.0 ± 3.5 | ≤ 1.0 | |
| Fhi (%) | 3.3 ± 1.2 | 2.9 ± 1.7 | 2.4 ± 1.2 | 0.9 ± 0.6 | 0.2 ± 0.3 | ≥ 2.0 | |
| Fhi$_{sca}$ (%) | 1.4 ± 1.7 | 0.9 ± 0.9 | 0.8 ± 0.8 | 0.3 ± 0.4 | 0.1 ± 0.1 | ≥ 1.0 | |
| Fhi$_{det}$ (%) | 11.5 ± 7.8 | 10.3 ± 5.7 | 6.7 ± 6 | 5.3 ± 5.1 | 1.8 ± 2.1 | ≥ 7.0 | |
| #runs combined | 26 | 9 | 15 | 8 | 9 | | |

**Notes.** The same as in Table 3.